\documentclass[aps,eqsecnum,preprint,floats,epsf,epsfig,nofootinbib,letter]{revtex4}
\usepackage{epsfig}
\usepackage{multirow}

\begin{document}
\def\be{\begin{eqnarray}}
\def\en{\end{eqnarray}}
\def\non{\nonumber}
\def\la{\langle}
\def\ra{\rangle}
\def\T{{\cal T}}
\def\O{{\cal O}}
\def\B{{\cal B}}
\def\ep{\varepsilon}
\def\hep{\hat{\varepsilon}}
\def\up{\uparrow}
\def\dw{\downarrow}
\def\ek{{\vec{\ep}_\perp\cdot\vec{k}_\perp}}
\def\epp{{\vec{\ep}_\perp\cdot\vec{P}_\perp}}
\def\kp{{\vec{k}_\perp\cdot\vec{P}_\perp}}
\def\lsim{ {\ \lower-1.2pt\vbox{\hbox{\rlap{$<$}\lower5pt\vbox{\hbox{$\sim$}
}}}\ } }
\def\gsim{ {\ \lower-1.2pt\vbox{\hbox{\rlap{$>$}\lower5pt\vbox{\hbox{$\sim$}
}}}\ } }
\def\dk{\partial\!\cdot\!K}
\def\pr{{\sl Phys. Rev.}~}
\def\prl{{\sl Phys. Rev. Lett.}~}
\def\pl{{\sl Phys. Lett.}~}
\def\np{{\sl Nucl. Phys.}~}
\def\zp{{\sl Z. Phys.}~}

\font\el=cmbx10 scaled \magstep2{\obeylines\hfill September, 2018}

\vskip 1.5 cm

\centerline{\large\bf Phenomenological Study of Heavy Hadron Lifetimes}
\bigskip
\centerline{\bf Hai-Yang Cheng}
\medskip
\centerline{ Institute of Physics, Academia Sinica}
\centerline{Taipei, Taiwan 115, Republic of China}
\medskip

\bigskip
\bigskip
\bigskip
\bigskip
\bigskip
\centerline{\bf Abstract}
\bigskip
\small
The lifetimes of bottom and charmed hadrons are analyzed within the framework of the heavy quark expansion (HQE). Lifetime differences arise from the spectator effects such as weak $W$-annihilation and Pauli interference. Spectator effects originating from dimension-7 four-quark operators are derived.
Hadronic matrix elements of four-quark operators are parameterized in a model-independent way.
Using the dimension-6 bag parameters recently determined from HQET sum rules and the vacuum insertion approximation for meson matrix elements of dimension-7 operators, the calculated $B$ meson lifetime ratios $\tau(B^+)/\tau(B^0_d)=1.074^{+0.017}_{-0.016}$ and $\tau(B^0_s)/\tau(B^0_d)=0.9962\pm0.0024$  are in excellent agreement with experiment. Likewise, based on the quark model evaluation of baryon matrix elements,
the resulting baryon lifetime ratios $\tau(\Xi_b^-)/\tau(\Lambda_b^0)$, $\tau(\Xi_b^-)/\tau(\Xi_b^0)$,  $\tau(\Omega_b^-)/\tau(\Xi_b^-)$ and the $\Lambda_b-B^0$ lifetime ratio $\tau(\Lambda_b^0)/\tau(B^0_d)=0.953$ also agree well with the data.
Contrary to the bottom hadron sector where the HQE in $1/m_b$ works well, the HQE to $1/m_c^3$ fails to give a satisfactory description of the lifetimes of both charmed mesons and charmed baryons. This calls for the subleading $1/m_Q$ corrections to spectator effects. The relevant dimension-7 spectator effects are in the right direction for explaining the large lifetime ratio of $\tau(\Xi_c^+)/\tau(\Lambda_c^+)$. However, the destructive $1/m_c$ corrections to $\Gamma(\Omega_c^0)$ are too large to justify the validity of the HQE, namely, the predicted Pauli interference and semileptonic rates for $\Omega_c^0$ become negative. Demanding these rates to be positive for a sensible HQE, we find the lifetime pattern
$\tau(\Xi_c^+)>\tau(\Omega_c^0)>\tau(\Lambda_c^+)>\tau(\Xi_c^0)$, contrary to the current hierarchy $\tau(\Xi_c^+)>\tau(\Lambda_c^+)>\tau(\Xi_c^0)>\tau(\Omega_c^0)$. We conclude that
the $\Omega_c^0$, which is naively expected to be shortest-lived in the charmed baryon system owing to the large constructive Pauli interference, could live longer than the $\Lambda_c^+$  due to the suppression from $1/m_c$ corrections arising from dimension-7 four-quark operators. The new charmed baryon lifetime pattern needs to be tested in forthcoming experiments.

\pagebreak


\section{Introduction}

It was realized since the late 1970s and 1980s that  the lifetime differences of singly heavy hadrons containing a heavy quark $Q$ arise mainly from
the spectator effects like $W$-exchange and Pauli interference
due to the identical quarks produced in heavy quark decay and in the wave function of the
heavy hadron \cite{Bigi92,BS93}. The spectator
effects were expressed in 1980s
in terms of local four-quark operators by relating the total widths to
the imaginary part of certain forward scattering amplitudes
\cite{Bilic,Guberina:1986,SV}.
With the advent of heavy quark effective theory (HQET), it was recognized
in early 1990s that nonperturbative corrections to the parton picture
can be systematically expanded in powers of $1/m_Q$ \cite{Bigi92,BS93}.
Within the QCD-based framework, namely the heavy
quark expansion (HQE), which is a generalization of the operator product
expansion (OPE) in $1/m_Q$ in the Minkowski space, some
phenomenological assumptions in 1980s acquired a firm theoretical footing in 1990s and nonperturbative effects can be systematically studied (for a review, see \cite{Lenz:2014}).

On the basis of the OPE approach for the analysis of inclusive weak
decays, the inclusive rate of the heavy hadron $H_Q$ is schematically
represented by
 \begin{eqnarray}
 \Gamma(H_Q\to f) = {G_F^2m_Q^5\over
192\pi^3}V_{\rm CKM}\left(A_0+{A_2\over m_Q^2}+{A_3\over
m_Q^3}+{A_4\over
m_Q^4}+{\cal O}\Big({1\over m_Q^5}\Big)\right),
 \end{eqnarray}
where $V_{\rm CKM}$ is the relevant CKM matrix element.
The $A_0$ term comes from the decay of the heavy quark $Q$ and is common to all
heavy hadrons $H_Q$. There is no linear $1/m_Q$ corrections to the
inclusive decay rate due to the lack of gauge-invariant
dimension-four operators \cite{Chay,Chay:1995,Bigi92}, a consequence known
as Luke's theorem \cite{Luke}. Nonperturbative corrections start
at order $1/m_Q^2$ and are model independent. Spectator
effects in inclusive decays due to the Pauli interference and
$W$-exchange contributions account for $1/m_Q^3$ corrections. The estimate of spectator
effects is model dependent; the hadronic four-quark matrix elements are usually
evaluated by assuming the factorization approximation for mesons and
the quark model for baryons. Moreover, there is a two-body phase-space enhancement factor of $16\pi^2$ for spectator effects relative to the three-body phase space for heavy quark decay. This means that $1/m_Q^3$ corrections can be quite significant. Moreover, spectator effects in charm hadron decays, being of
order $1/m_c^3$, can be comparable to and even exceed the $A_0$
term.

   Based on the HQE approach for the analysis of inclusive weak decays,
the first correction to bottom hadron lifetimes is
of order $1/m_b^2$ and it is model independent. For example, it was found in \cite{Neubert97} that
${\tau(B^-)/\tau(B_d)}\approx {\tau(B_s)/\tau(B_d)}= 1+{\cal O}(1/m_b^3)$
and ${\tau(\Lambda_b)/\tau(B_d)}= 0.98+{\cal O}(1/m_b^3)$. The $1/m_b^3$ corrections can be expressed in a model-independent manner \cite{Neubert97}
\be \label{eq:ratio3}
{\tau(\Lambda_b)\over\tau(B_d)} &\simeq &0.98-0.17\ep_1+0.20\ep_2-(0.012+
0.021\tilde B)r,
\en
where $\ep_i,~B_i,~\tilde B,~r$ are the hadronic parameters to be introduced
below in Sec.~III.A.
Experimentally, the $\Lambda_b^0$ lifetime was significantly shorter than the $B$ meson one in the early measurements. Taking the current $B^0$ meson lifetime $\tau(B^0)=(1.520\pm 0.004)$ ps \cite{PDG} as a benchmark, $\tau(\Lambda_b)$ was found to be $(1.14\pm0.08)$ ps in 1996 \cite{PDG1996}.  The world-averaged lifetime ratio then was
\be  \label{eq:ratio1996}
{\tau(\Lambda_b^0)\over \tau(B^0)}=0.79\pm0.06  \qquad (1996),
\en
dominated by CERN $e^+e^-$ collider LEP experiments \cite{LEP}. This lifetime ratio remained essentially unchanged even in 2004 \cite{HFAG:2004}
\be \label{eq:ratio2004}
{\tau(\Lambda_b^0)\over \tau(B^0)}=0.803\pm0.047\,,  \qquad {\rm HFAG~2004}.
\en
Since the two parameters $\ep_1$ and $\ep_2$ obey the constraint $\ep_1\approx 0.3\ep_2$ \cite{Neubert97} and they vanish under vacuum insertion approximation, it is very difficult to employ the HQE prediction  (\ref{eq:ratio3}) to accommodate the experimental value of $\tau(\Lambda_b)/\tau(B_d)$ without invoking too large a value of $r$ and/or $\tilde B$.
It is thus reasonable to conclude that the $1/m_b^3$ corrections
in the HQE do not suffice to describe the observed lifetime
difference between $\Lambda_b$ and $B_d$.

Motivated by the conflict between theory and experiment for the lifetime
ratio $\tau(\Lambda_b)/\tau(B_d)$, it was suggested in \cite{Altarelli} that
the assumption of local duality is not correct for nonleptonic inclusive
width and that the presence of linear $1/m_b$ corrections prohibited in the HQE is strongly
implied by the data. As shown in \cite{Cheng:1997} and \cite{Ito}, the simple ansatz of replacing $\Gamma_{\rm NL}$ by
$\Gamma_{\rm NL}(m_{\Lambda_b}/m_b)^5$  not only solves the lifetime ratio problem
but also provides the correct absolute decay widths for the $\Lambda_b$
baryon and the $B$ meson. However, there exist several insurmountable difficulties with this scenario and some of them were already discussed in \cite{Cheng:1997}.

Nowadays we know that the issue with the low value of $\Lambda_b-B^0$ lifetime ratio or the short $\Lambda_b$ lifetime was on the experimental side.
The first direct measurement of the lifetime ratio $\tau(\Lambda_b^0)/\tau(B^0)$ was carried out by the D0 Collaboration in 2005 with the result $0.87^{+0.17}_{-0.14}\pm0.03$ \cite{D0}. Also, the CDF experiment measured the $\Lambda_b$ lifetime in exclusive decay to $J/\psi \Lambda$ \cite{CDF} and showed that the $\Lambda_b$ lifetime is significantly longer than either previous $\Lambda_b$ lifetime measures or state-of-the-art calculation based on the HQE at the time.
The world averages as of today are \cite{PDG}
\be
\tau(\Lambda_b^0)=(1.470\pm0.010)\times 10^{-12}\,s\,, \qquad \tau(\Lambda_b^0)/\tau(B^0_d)=0.964\pm0.007\,.
\en
As we shall see in Sec. IV.A below, the current value of the $\Lambda_b-B^0$ lifetime ratio can be explained nicely in the HQE approach.

The major theoretical uncertainties of the HQE predictions for hadron lifetimes come from the hadronic matrix elements of four-quark operators. In the meson sector, the meson matrix elements can be expressed in a model-independent manner in terms of four bag parameters $B_{1,2}$ and $\epsilon_{1,2}$. These parameters have been calculated using lattice QCD and QCD sum rules (see \cite{Lenz:2014} for a review). Based on HQET sum rules, they have been updated recently in \cite{Kirk}. For the baryon matrix elements, they can be parameterized in terms of four parameters $L_{1,\cdots,4}$, but only two of them are independent.

Although the heavy quark expansion in $1/m_b$ works well for $B$ mesons and bottom baryons, the HQE in $1/m_c$ fails to give a satisfactory description of the lifetimes of both charmed mesons and charmed baryons. First of all, to order $1/m_c^3$, the destructive Pauli interference in $D^+$ decay overcomes the $c$ quark decay rate so that the inclusive rate and the lifetime of $D^+$ become negative. Hence, it is not meaningful to discuss the lifetime ratio $\tau(D^+)/\tau(D^0)$ at this level. Second, the lifetime pattern of charmed baryons is understandable only qualitatively, but not quantitatively. The quantitative estimates of charmed baryon lifetimes and their ratios are still rather poor \cite{Cheng:1997}. For example, $\tau(\Xi_c^+)/\tau(\Lambda_c^+)$ is calculated to be 1.03 (see Sec. IV.B below), while experimentally it is measured to be $2.21\pm0.15$ \cite{PDG}. Therefore, it is natural to consider the effects stemming from the next-order $1/m_c$ expansion. Specifically, we shall consider the subleading $1/m_c$ corrections to the spectator effects.

The $1/m_Q$ corrections to the spectator effects are computed by expanding the
forward scattering amplitude in the light-quark momentum and matching the result onto the operators containing derivative insertions. Dimension-7 terms are either the four-quark operators times the spectator quark mass or the four-quark operators with an additional derivative \cite{Gabbiani:2003pq,Gabbiani:2004tp}.
Dimension-7 operators were first studied in \cite{Beneke:1996gn} for the width difference in the $B_s-\bar B_s$ system, in
\cite{Gabbiani:2003pq,Gabbiani:2004tp} for the lifetime differences of heavy hadrons and in \cite{Lenz:D} for $D$-meson lifetimes.

In this work we will study spectator effects in inclusive
nonleptonic and semileptonic decays,
analyze the lifetime pattern of heavy hadrons, and pay attention to the effects of dimension-7 operators on the heavy hadron lifetimes, especially for the charmed mesons and baryons. Our goal is to see if the aforementioned problems such as the negative lifetime of the $D^+$ meson and the closeness of $\Xi_c^+$ and $\Lambda_c^+$ lifetimes can be resolved by the inclusion of subleading $1/m_Q$ corrections to the spectator effects.

This work is organized as follows. In Sec.~II we give
general heavy quark expansion expressions for inclusive nonleptonic and
semileptonic widths. We then study lifetimes of $B$ and $D$ mesons
in Sec.~III and bottom and charmed baryons in Sec.~IV with the evaluation of hadronic four-quark matrix elements.
Discussions and conclusions are given in Sec.~V. In Appendix A we  sketch the derivation of dimension-7 four-quark operators relevant for the spectator effects in heavy baryon decays. Appendix B is devoted to the evaluation of baryon matrix elements in the quark model.

\section{Framework}
   In this section we write down the general expressions for the
inclusive decay widths of heavy hadrons and leave the evaluation of the relevant hadronic matrix elements to the next section. It is known that the inclusive decay rate is governed by
the imaginary part of an effective nonlocal forward transition operator $T$.
When the energy released in the decay is large enough, the nonlocal effective
action can be recast as an infinite series of local operators with
coefficients containing inverse powers of the heavy quark mass $m_Q$.
Under this heavy quark expansion, the inclusive
nonleptonic decay rate of a singly heavy hadron $H_Q$ containing a heavy quark $Q$
is given by \cite{Bigi92,BS93}

\be
\Gamma(H_Q)={1\over 2m_{H_Q}}{\rm Im}\,\la H_Q|T|H_Q\ra={1\over 2m_{H_Q}}\la H_Q|\int d^4x\,T[{\cal L}^\dagger_W(x){\cal L}_W(0)]|H_Q\ra,
\en
where the second $T$ appearing in the integral is a time-ordering symbol.
Under the operator product expansion, the transition operator $T$ can be expressed in terms of local quark operators
\be
{\rm Im}\,T={G_F^2m_Q^5\over 192\pi^3}\,\xi\,\left(c_{3,Q}\bar QQ+{c_{5,Q}\over m_Q^2}\bar Q\sigma\cdot G Q+{c_{6,Q}\over m_Q^3}T_6+ {c_{7,Q}\over m_Q^4}T_7+\cdots\right),
\en
where $\xi$ is the relevant CKM matrix element,
the dimension-6 $T_6$ consists of the four-quark operators $(\bar Q\Gamma q)(\bar q\Gamma Q)$ with $\Gamma$ representing a combination of the Lorentz and color matrices, while a subset of dimension-7 $T_7$ is governed by the four-quark operators containing derivative insertions (see Sec.II.B below). Since $\sigma\cdot G=-2\vec{\sigma}\cdot\vec{B}$, the $\bar Q\sigma\cdot G Q$ term describes the interaction of the heavy $Q$ quark spin with the gluon field. Explicitly,
\be \label{eq:NLrate}
\Gamma_{\rm NL}(H_Q) &=& {G_F^2m_Q^5\over 192\pi^3}\,\xi\,
\Bigg\{ c_{3,Q}^{\rm NL} \Big[1-{\mu_\pi^2-\mu^2_G\over 2m_Q^2}\Big]+ 2c_{5,Q}^{\rm NL} {\mu^2_G\over m_Q^2} \non \\ &+& {c_{6,Q}^{\rm NL} \over m_Q^3}{\la H_Q|T_6|H_Q\ra\over 2m_{H_Q}}+{c_{7,Q}^{\rm NL} \over m_Q^4}{\la H_Q|T_7|H_Q\ra\over 2m_{H_Q}}+\cdots\Bigg\},
\en
where use of
\be
{\la H_Q|\bar QQ|H_Q\ra\over 2m_{H_Q}}=1-{\mu_\pi^2\over 2m_Q^2}+{\mu_G^2\over 2m_Q^2}
\en
has been made with
\be \label{eq:lambda12}
&& \mu_\pi^2\equiv {1\over 2m_{H_Q}}\la H_Q|\bar Q(i\vec{D})^2Q|H_Q\ra=-{1\over 2m_{H_Q}}\la H_Q|\bar Q(i{D_\perp})^2Q|H_Q\ra=
-\lambda_1,   \non \\
&& \mu_G^2\equiv{1\over 2m_{H_Q}}\la H_Q|\bar Q{1\over 2}\sigma\cdot
GQ|H_Q\ra=d_H\lambda_2.
\en

In heavy quark effective theory, the mass of the heavy hadron $H_Q$ is  of the form
\be \label{eq:mHQ}
m_{H_Q}=\,m_Q+\bar\Lambda_{H_Q}-{\lambda_1\over 2m_Q}-{d_H\lambda_2\over 2m_Q},
\en
where the three nonperturbative HQET parameters $\bar\Lambda_{H_Q},~
\lambda_1$ and $\lambda_2$ are independent of the heavy quark mass and
$\bar\Lambda_{H_Q}$ can be regarded as the binding energy of the heavy hadron in the infinite mass limit \cite{Neubert96}. Since the chromomagnetic field is produced by the light cloud
inside the heavy hadron, it is clear that $\sigma\cdot G$ is proportional to
$\vec{S}_Q\cdot\vec{S}_\ell$,  where $\vec{S}_Q~(\vec{S}_\ell)$ is the spin
operator of the heavy quark (light cloud). The parameter $d_H$ is given by
\be
d_H &=& -\la H_Q|4\vec{S}_Q\cdot\vec{S}_\ell|H_Q\ra   \non \\
&=& -2[S_{\rm tot}(S_{\rm tot}+1)-S_Q(S_Q+1)-S_\ell(S_\ell+1)].
\en
Therefore, $d_H=3$ for $B$ and $D$ mesons, $d_H=-1$ for $B^*$ and $D^*$ mesons, $d_H=0$
for the antitriplet baryon $T_Q$, $d_H=4$ for the spin-${1\over 2}$
sextet baryon $S_Q$ and $d_H=-2$ for the spin-${3\over 2}$ sextet baryon
$S^*_Q$. It follows from Eq. (\ref{eq:mHQ}) that
\be
\lambda_2^{\rm meson} &=& {1\over 4}(m^2_{P^*}-m_P^2)=\cases{ 0.12\,{\rm
GeV}^2& for~$B$~meson,   \cr  0.14\,{\rm GeV}^2 & for~$D$~meson,   \cr}   \non \\
\lambda_2^{\rm baryon} &=& {1\over 6}(m^2_{S^*_Q}-m^2_{S_Q}).
\en
Numerically (in units of GeV$^2$),
\be
&& \lambda_2^{\Sigma_c}=0.054, \qquad \lambda_2^{\Xi_c'}=0.061, \qquad \lambda_2^{\Omega_c}=0.064, \non \\
&& \lambda_2^{\Sigma_b}=0.040, \qquad \lambda_2^{\Xi_b'}=0.040, \qquad \lambda_2^{\Omega_b}=0.041\,.
\en
It is interesting to note that the large-$N_c$ relation \cite{Jenkins,Jenkins:1997}
\be
\lambda_2^{\rm meson}\sim N_c\lambda_2^{\rm baryon}
\en
is fairly satisfied especially for bottom hadrons.
As for the kinetic energy parameter $\lambda_1$,  we shall use
\cite{Gambino:2016}
\be
\lambda_1^{\rm meson}\sim \lambda_1^{\rm baryon}=-(0.432\pm 0.068)\,{\rm GeV}^2.
\en

Summing over the contributions from $b\to c\bar cs$, $b\to c\bar cd$, $b\to c\bar ud$ and $b\to c\bar us$ processes, we have \cite{Bigi92,BS93}
\be \label{eq:c3,5b}
c_{3,b}^{\rm NL}  &=& \left(N_cc_1^2+N_cc_2^2+2c_1c_2\right)\left(I_0(x,0,0)+I_0
(x,x,0)\right), \non \\
c_{5,b}^{\rm NL}  &=& - \left(N_cc_1^2+N_cc_2^2+2c_1c_2\right)(I_1(x,0,0)+I_1(x,x,0)) \non \\
 && -8c_1c_2(I_2(x,0,0)+I_2(x,x,0)),
\en
where $x=(m_c/m_b)^2$ and the good approximations $|V_{ud}|^2+|V_{us}|^2\approx 1$, $|V_{cs}|^2+|V_{cd}|^2\approx 1$ have been made.
In the above equation,  $c_1,~c_2$ are Wilson
coefficient functions, $N_c=3$ is the number of color,
$I_0,~I_1$ and $I_2$ are the phase-space factors:
\be
I_0(x,0,0) &=& (1-x^2)(1-8x+x^2)-12x^2\ln x,   \non \\
I_1(x,0,0) &=& {1\over 2}(2-x{d\over dx})I_0(x,0,0)=(1-x)^4,   \non \\
I_2(x,0,0) &=& (1-x)^3,
\en
for $b\to c\bar ud$ and $b\to c\bar us$ ($x=m_c^2/m_b^2$) or $c\to su\bar d$ ($x=m_s^2/m_c^2$)
transition and
\be
I_0(x,x,0) &=& v(1-14x-2x^2-12x^3)+24x^2(1-x^2)\ln {1+v\over 1-v},   \non \\
I_1(x,x,0) &=& {1\over 2}(2-x{d\over dx})I_0(x,x,0)=v(1-2x)(1-4x-6x^2)+24x^4\ln {1+v\over 1-v},   \non \\
I_2(x,x,0) &=& v(1+{x\over 2}+3x^2)-3x(1-2x^2)\ln {1+v\over 1-v},
\en
for $b\to c\bar c  s$ or $c\to s\bar su$ transition with $v\equiv\sqrt{1-4x}$.
For the $c$ quark decay, contributions from $c\to s\bar du$ and $c\to s\bar su$ yield
\be
c_{3,c}^{\rm NL} &=& \left(N_cc_1^2+N_cc_2^2+2c_1c_2\right)\left(I_0(x,0,0)|V_{ud}|^2+I_0
(x,x,0)|V_{us}|^2\right),  \non \\
c_{5,c}^{\rm NL} &=& - \left(N_cc_1^2+N_cc_2^2+2c_1c_2\right)\left(I_1(x,0,0)|V_{ud}|^2+I_1(x,x,0)|V_{us}|^2\right) \\
&& - 8c_1c_2\left(I_2(x,0,0)|V_{ud}|^2+I_2(x,x,0)|V_{us}|^2\right), \non
\en
with $x=(m_s/m_c)^2$.

It is now ready to deduce
the inclusive semileptonic widths from Eq. (\ref{eq:NLrate}) by putting $c_1=1$, $c_2=0$ and $N_c=1$:
\be \label{eq:SLrate}
\Gamma_{\rm SL}(H_Q) &=& {G_F^2m_Q^5\over 192\pi^3}\,\xi\,
\Bigg\{ c_{3,Q}^{\rm SL} \Big[1-{\mu_\pi^2-\mu^2_G\over 2m_Q^2}\Big]+ 2c_{5,Q}^{\rm SL}{\mu^2_G\over m_Q^2} \Bigg\},
\en
where
\be \label{eq:c3,5bSL}
c_{3,b}^{\rm SL}(x,x_\tau) &=& 2I_0(x,0,0)+I_0
(x,x_\tau,0),  \non \\
c_{5,b}^{\rm SL}(x,x_\tau) &=& -\left(2I_1(x,0,0)+I_1(x,x_\tau,0)\right),
\en
and
\be \label{eq:c3,5cSL}
c_{3,c}^{\rm SL}(x,x_\mu) &=& I_0(x,0,0)+I_0
(x,x_\mu,0),  \non \\
c_{5,c}^{\rm SL}(x,x_\mu) &=& -\left(I_1(x,0,0)+I_1(x,x_\mu,0)\right),
\en
with $x_\ell=(m_\ell/m_Q)^2$.  For the expression of $I_{0,1}(x,y,0)$ with $y\neq x$, see \cite{Falk:1994} or the appendix of \cite{Mannel} with $C_0=I_0(x,y,0)$ and $C_{\mu^2_G}=I_1(x,y,0)$.

\subsection{Dimension-6 operators}
\begin{figure}[t]
\begin{center}
\includegraphics[height=19mm]{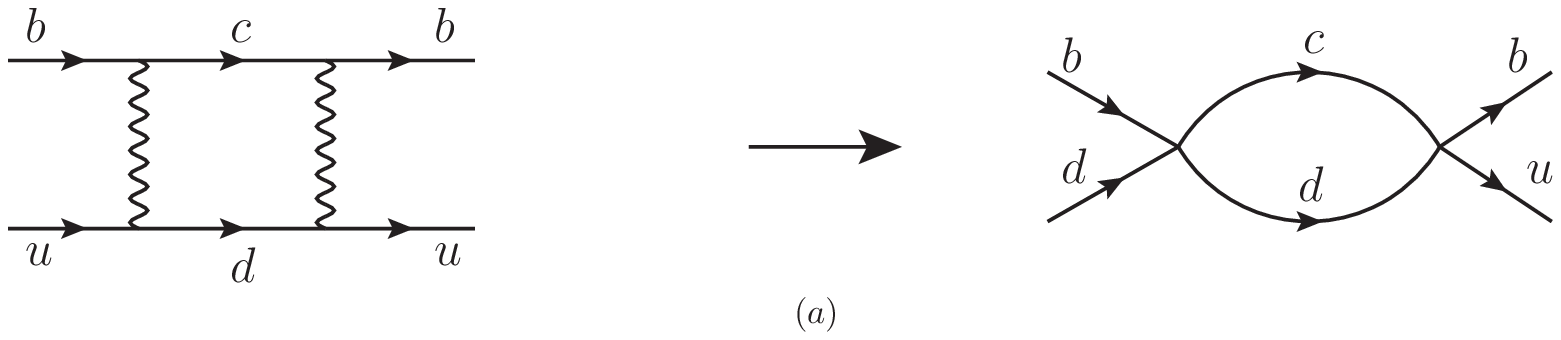}
\includegraphics[height=27mm]{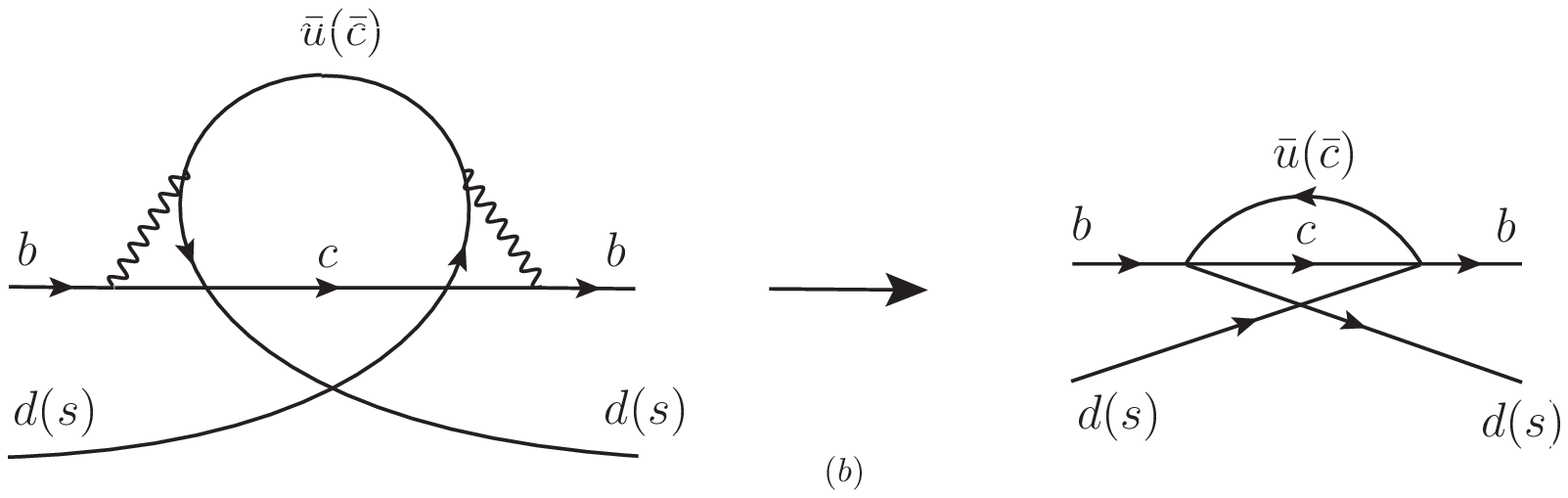}
\includegraphics[height=32mm]{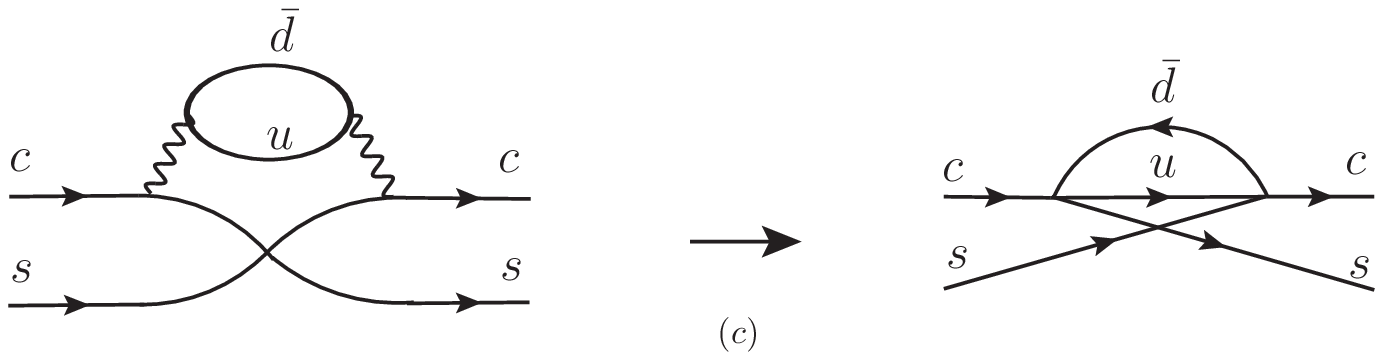}
\caption{Spectator effects in heavy baryon decays: (a) $W$-exchange, (b) destructive Pauli interference for $b\to c\bar ud$ and $b\to c\bar cs$, and (c) constructive Pauli interference for $c\to u\bar ds$.}
\label{fig:spectator}
\end{center}
\end{figure}

 Defining
\be
\T_6={G_F^2m_Q^2\over 192\pi^3}\xi\,c_{6,Q}^{\rm NL}\,T_6,
\en
the dimension-six four-quark operators in Eq. (\ref{eq:NLrate}) responsible for spectator effects in inclusive decays of heavy baryons (denoted by $\B_Q$) are given by
\cite{Bilic,Guberina:1986,SV}
\be  \label{eq:T6baryon}
\T_{6,ann}^{\B_Q,q_1} &=& {G^2_Fm_Q^2\over 2\pi}\,\xi\,(1-x)^2\Big\{
(c_1^2+c_2^2)(\bar Q Q)(\bar q_1q_1)+2c_1c_2(\bar Qq_1)(\bar q_1 Q)\Big\},
 \non \\
\T_{6,int-}^{\B_Q,q_2} &=& -{G_F^2m_Q^2\over 6\pi}\,\xi (1-x)^2\Bigg\{ c_1^2\left[ (1+
{x\over 2})(\bar QQ)(\bar q_2q_2)-(1+2x)\bar Q^\alpha(1-\gamma_5)q_2^\beta
\bar q_2^\beta(1+\gamma_5)Q^\alpha\right]   \non \\
&+& (2c_1c_2+N_cc_2^2)\left[ (1+{x\over 2})(\bar Qq_2)(\bar q_2Q)-
(1+2x)\bar Q(1-\gamma_5)q_2\bar q_2(1+\gamma_5)Q\right]\Bigg\},  \non \\
\T_{6,int-}^{\B_Q,q_3} &=& - {G_F^2m_Q^2\over 6\pi}\,\xi\sqrt{1-4x}\,\Bigg\{
c_1^2\,\Big[ (1-x)(\bar Q Q)(\bar q_3q_3)-(1+2x)\bar Q^\alpha(1-
\gamma_5)q_3^\beta\bar q_3^\beta (1+\gamma_5)Q^\alpha\Big]   \non \\
&+& (2c_1c_2+N_cc_2^2)\,\Big[ (1-x)(\bar Qq_3)(\bar q_3Q)-
(1+2x)\bar Q(1-\gamma_5)q_3\bar q_3(1+\gamma_5)Q\Big]\Bigg\}, \non\\
\T_{6,int+}^{\B_Q,q_3} &=& -{G_F^2m_Q^2\over 6\pi}\,\xi\Bigg\{
c_2^2\left[(\bar QQ)(\bar q_3q_3)-\bar Q^\alpha(1-\gamma_5)q_3^\beta
\bar q_3^\beta(1+\gamma_5)Q^\alpha\right]   \non \\
&+& (2c_1c_2+N_cc_1^2)\Big[ (\bar Qq_3)(\bar q_3Q)-
\bar Q(1-\gamma_5)q_3\bar q_3(1+\gamma_5)Q\Big] \Bigg\},
\en
where $(\bar q_1q_2)\equiv \bar q_1\gamma_\mu(1-\gamma_5)q_2$, $\alpha,~\beta$
are color indices and $\xi$ is the relevant CKM matrix element for the quark-mixing-favored decay.
Note that for charm decay, $Q=c,~q_1=d,~q_2=u$ and $q_3=s$ and for bottom
decay, $Q=b,~q_1=u,~q_2=d,~q_3=s$. In the baryon sector, the first term $\T_{6,ann}^{\B_Q,q_1}$ corresponds to a $W$-exchange (or generically weak annihilation) contribution (see Fig. \ref{fig:spectator}(a)), the rest to contributions from Pauli interference. For example, $\T_{6,int-}^{\B_Q,q_2}$ arises from the destructive interference of the $q_2$ quark produced in the heavy quark $Q$ decay with the $q_2$ quark in the wave function of the heavy baryon $\B_Q$ (Fig. \ref{fig:spectator}(b)). The last term $\T_{6,int+}^{\B_Q,q_3}$ in (\ref{eq:T6baryon}) is due to the constructive interference of the $s$ quark and hence it occurs only
in charmed baryon decays, i.e. $Q=c$ and $q_3=s$ (Fig. \ref{fig:spectator}(c)). The third term $\T_{6,int-}^{\B_Q,q_3}$ comes from the destructive Pauli interference due to $b\to c\bar cs$ (Fig. \ref{fig:spectator}(b)) or $c\to s\bar su$.
This term  exists  in bottom decays with $c\bar c$ intermediate
states and in charm decays with $s\bar s$ intermediate
states.

As we shall see in Sec. IV, Pauli interferences described by $\T_{6,int-}^{\B_Q,q_2}$ and $\T_{6,int-}^{\B_Q,q_3}$ are destructive as the relevant Wilson coefficient of the form $(N_cc_2^2+2c_1c_2-\tilde Bc_1^2)$ with the hadronic parameter $\tilde B$ defined in Eq. (\ref{eq:Btilde}) is negative in both charm and bottom sectors, whereas the Pauli interference from $\T_{6,int+}^{\B_Q,q_3}$ is constructive as the relevant Wilson coefficient $(N_cc_1^2+2c_1c_2-\tilde Bc_2^2)$ is positive. This is not necessarily true for dimension-7 Pauli interference effects to be described below.

In the heavy meson sector,
the $W$-exchange contribution to the heavy meson corresponds to the Pauli interference term $\T_{6,int-}^{\B_Q}$ in Eq. (\ref{eq:T6baryon}) in heavy baryon decays, while
the Pauli interference in inclusive nonleptonic decays of heavy mesons corresponds to the annihilation term $\T_{6,ann}^{\B_Q}$ in heavy baryon decays. This will be discussed in Sec. III.

It is clear from Eqs.~(\ref{eq:NLrate}) and (\ref{eq:T6baryon}) that there is a two-body
phase-space enhancement factor of $16\pi^2$ for spectator effects
relative to the three-body phase space for the heavy quark decay. This implies
that spectator effects, being of order $1/m_Q^3$, are comparable to and
even exceed the $1/m_Q^2$ terms. Note that the Wilson coefficients and four-quark operators in Eq.~(\ref{eq:T6baryon}) are
renormalized at the heavy quark mass scale. Sometimes the so-called hybrid
renormalization \cite{SV,SV87} is performed to evolve the four-quark operators
(not the Wilson coefficients!) from $m_Q$ down to a low-energy scale, say, a
typical hadronic scale $\mu_{\rm had}$.  The evolution from $m_Q$ down to $\mu_{\rm had}$ will in
general introduce new structures such as penguin operators. Nevertheless, in
the present paper we will follow \cite{Neubert97} to employ (\ref{eq:NLrate}) and (\ref{eq:T6baryon}) as our starting point for describing inclusive weak decays
since it is equivalent to first evaluating the four-quark
matrix elements renormalized at the $m_Q$ scale and then relating them
to the hadronic matrix elements renormalized at $\mu_{\rm had}$ through
the renormalization group equation, provided that the effect of penguin
operators is neglected.

\begin{figure}[t]
\begin{center}
\includegraphics[height=30mm]{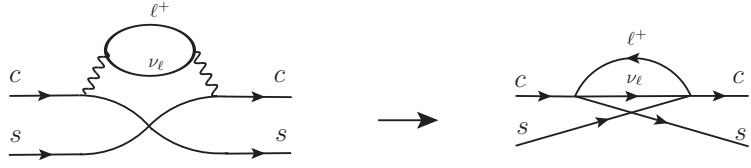}
\caption{Spectator effect in the charmed-baryon
semileptonic decay.}
\label{fig:CharmSL}
\end{center}
\end{figure}

   For inclusive semileptonic decays, apart from the heavy quark decay
contribution there is an additional spectator effect in charmed-baryon
semileptonic decay originating from the Pauli interference of the $s$
quark \cite{Voloshin}; that is, the $s$ quark produced in $c\to s\ell^+\nu_\ell$ has an interference with the $s$ quark in the wave function of the charmed baryon (see Fig. \ref{fig:CharmSL}).
It is now ready to deduce this term from $\T_{6,int+}^{q_3}$ in Eq. (\ref{eq:T6baryon})  by putting $c_1=1$, $c_2=0$, $N_c=1$ and $q_3=s$:
\be \label{eq:SL6}
\Gamma^{\rm SL}_{6,int}(\B_c) =
 -{G_F^2m_c^2\over 6\pi}|V_{cs}|^2{1\over 2m_{\B_c}}  \la \B_c|
(\bar c s)(\bar s c)-\bar c(1-\gamma_5)s\bar s(1+\gamma_5)c |\B_c\ra.
\en
Obviously, this term occurs only in the semileptonic decays of $\Xi_c$ and $\Omega_c$ baryons.

\subsection{Dimension-7 operators}
To the order of $1/m_Q^4$ in the heavy quark expansion in Eq. (\ref{eq:NLrate}), we need to consider dimension-7 operators. For our purposes, we shall focus on the $1/m_Q$
corrections to the spectator effects discussed in the last subsection and neglect the operators with gluon fields. As mentioned in the Introduction,
the relevant dimension-7 terms are either the four-quark operators times the spectator quark mass or the four-quark operators with an additional derivative \cite{Gabbiani:2003pq,Gabbiani:2004tp}.
We shall follow \cite{Lenz:D} to define the following dimension-7 four-quark operators:
\be
&& P_1^q={m_q\over m_Q}\bar Q(1-\gamma_5)q\bar q(1-\gamma_5)Q, \qquad\qquad\qquad
~~P_2^q={m_q\over m_Q}\bar Q(1+\gamma_5)q\bar q(1+\gamma_5)Q, \non \\
&& P_3^q={1\over m_Q^2}\bar Q \stackrel{\leftarrow}{D}_\rho\gamma_\mu(1-\gamma_5)D^\rho q\bar q\gamma^\mu(1-\gamma_5)Q, \quad
~P_4^q={1\over m_Q^2}\bar Q \stackrel{\leftarrow}{D}_\rho(1-\gamma_5)D^\rho q\bar q(1+\gamma_5)Q, \non \\
&& P_5^q={1\over m_Q}\bar Q \gamma_\mu(1-\gamma_5)q\bar q\gamma^\mu(1-\gamma_5)(iD\!\!\!\!/)Q, \quad\quad
~~P_6^q={1\over m_Q}\bar Q (1-\gamma_5)q\bar q(1+\gamma_5)(iD\!\!\!\!/)Q,
\en
and the color-octet operators $S_i^q$ ($i=1,...,6$) obtained from $P_i^q$ by inserting $t^a$ in the two currents of the respective color singlet operators, for example, $S_1^q={m_q\over m_Q}\bar Q(1-\gamma_5)t^aq\bar q(1-\gamma_5)t^a Q$.

Following the prescription outlined in \cite{Lenz:D}, one can derive the dimension-7 operators relevant to heavy baryon decays.
Explicitly (see Appendix A for details),
\be \label{eq:T7baryon_original}
\T_{7,ann}^{\B_Q,q_1} &=& {G^2_Fm_Q^2\over 6\pi}\,\xi\,(1-x)\Bigg\{ \left(2N_c c_1c_2+c_1^2+c_2^2\right)\Big[2(1+x)P_3^{q_1}+(1-x)P^{q_1}_5\Big] \non \\
&+& 6(c_1^2+c_2^2)\Big[2(1+x)S_3^{q_1}+(1-x)S^{q_1}_5\Big]  \Bigg\},   \non \\
\T_{7,int}^{\B_Q,q_2} &=& {G_F^2m_Q^2\over 6\pi}\,\xi (1-x)\Bigg\{  \left({1\over N_c}c_1^2+2c_1c_2+N_cc_2^2\right)\Big[ -(1-x)(1+2x)(P_1^{q_2}+P_2^{q_2}) \non \\
&+& 2(1+x+x^2)P_3^{q_2}-12x^2P_4^{q_2} -(1-x)(1+{x\over 2})P_5^{q_2}+(1-x)(1+2x)P_6^{q_2}\Big] \non \\
&+& 2c_1^2 \Big[ -(1-x)(1+2x)(S_1^{q_2}+S_2^{q_2})+2(1+x+x^2)S_3^{q_2}-12x^2S_4^{q_2}  \non \\
&-& (1-x)(1+{x\over 2})S_5^{q_2}+(1-x)(1+2x)S_6^{q_2}\Big]
\Bigg\},    \\
\T_{7,int}^{\B_Q,q_3} &=& {G_F^2m_Q^2\over 6\pi}\,\xi \sqrt{1-4x}\Bigg\{  \left({1\over N_c}c_1^2+2c_1c_2+N_cc_2^2\right)\Big[ -(1+2x)(P_1^{q_3}+P_2^{q_3}) \non \\
&+&{2\over 1-4x}(1-2x-2x^2)P_3^{q_3} - {24x^2\over 1-4x}P_4^{q_3}-(1-x)P_5^{q_3}+(1+2x)P_6^{q_3}\Big] \non \\
&+& 2c_1^2 \Big[ -(1+2x)(S_1^{q_3}+S_2^{q_3})+{2\over 1-4x}(1-2x-2x^2)S_3^{q_3}
-{24x^2\over 1-4x}S_4^{q_3} \non \\
&-& (1-x)S_5^{q_3}+(1+2x)S_6^{q_3} \Big]\Bigg\}, \non \\
\T_{7,int}^{\B_c,s} &=& {G_F^2m_c^2\over 6\pi}\,\xi \Bigg\{  \left({1\over N_c}c_2^2+2c_1c_2+N_cc_1^2\right)\Big[ -P_1^s-P_2^s +2P_3^s-P_5^s+P_6^s \Big] \non \\
&+& 2c_2^2 \Big[ -S_1^s-S_2^s+2S_3^s-S^s_5+S_6^s \Big]
 \Bigg\}. \non
\en
However, we shall see later that in order to evaluate the baryon matrix elements, it is more convenient to express dimension-7 operators in terms of $P_i^q$ and $\tilde P_i^q$ ones, where $\tilde P_i$ denotes the color-rearranged operator that follows from the expression of $P_i$ by interchanging the color indices of the $q_i$ and $\bar q_j$ Dirac spinors. For example, $\tilde P_1^q={m_q\over m_Q}\bar Q_i(1-\gamma_5)q_j\bar q_i(1-\gamma_5)Q_j$. Using the relation
\be
S_i=-{1\over 2N_c}P_i+{1\over 2}\tilde P_i,
\en
we obtain
\be \label{eq:T7baryon}
\T_{7,ann}^{\B_Q,q_1} &=& {G^2_Fm_Q^2\over 2\pi}\,\xi\,(1-x)\Bigg\{ 2c_1c_2\Big[2(1+x)P_3^{q_1}+(1-x)P^{q_1}_5\Big] \non \\
&+& (c_1^2+c_2^2)\left[2(1+x)\tilde P_3^{q_1}+(1-x)\tilde P^{q_1}_5\right]\Bigg\},
\non\\
\T_{7,int}^{\B_Q,q_2} &=& {G_F^2m_Q^2\over 6\pi}\,\xi(1-x)\Bigg\{ \Big(2c_1c_2+N_cc_2^2\Big)\Big[-(1-x)(1+2x)(P_1^{q_2}+P_2^{q_2}) \non \\
&+& 2(1+x+x^2)P_3^{q_2}
-12x^2P_4^{q_2}-(1-x)(1+{x\over 2})P_5^{q_2}+(1-x)(1+2x)P_6^{q_2}\Big] \non \\
 &+& c_1^2\Big[ -(1-x)(1+2x)(\tilde P_1^{q_2}+\tilde P_2^{q_2})
+ 2(1+x+x^2)\tilde P_3^{q_2}-12x^2\tilde P_4^{q_2} \non \\
&-& (1-x)(1+{x\over 2})\tilde P_5^{q_2}+(1-x)(1+2x)\tilde P_6^{q_2}\Big]
\Bigg\},   \\
\T_{7,int}^{\B_Q,q_3} &=& {G_F^2m_Q^2\over 6\pi}\,\xi \sqrt{1-4x}\Bigg\{  \left(2c_1c_2+N_cc_2^2\right)\Big[ -(1+2x)(P_1^{q_3}+P_2^{q_3}) \non \\
&+& {2\over 1-4x}(1-2x-2x^2)P_3^{q_3} - {24x^2\over 1-4x}P_4^{q_3} -(1-x)P_5^{q_3}+(1+2x)P_6^{q_3}\Big] \non \\
&+& c_1^2 \Big[ -(1+2x)(\tilde P_1^{q_3}+\tilde P_2^{q_3})
+{2\over 1-4x}(1-2x-2x^2)\tilde P_3^{q_3}
-{24x^2\over 1-4x}\tilde P_4^{q_3} \non \\
&-&(1-x)\tilde P_5^{q_3}+(1+2x)\tilde P_6^{q_3}
\Big] \Bigg\}, \non \\
\T_{7,int}^{\B_c,s} &=& {G_F^2m_c^2\over 6\pi}\,\xi\Bigg\{ \left(2c_1c_2+N_cc_1^2\right)\Big[-P_1^s-P_2^s
+ 2P_3^s -P_5^s+P_6^s\Big] \non \\
 &+& c_2^2\Big[ -\tilde P_1^s-\tilde P_2^s
+ 2\tilde P_3^s - \tilde P_5^s+\tilde P_6^s\Big]
\Bigg\}. \non
\en

For the dimension-7 operators relevant to heavy meson decays, see the next section.

\subsection{Lifetime ratio}
In order to compare the HQE predictions with the experimental results, we often consider the lifetime ratio of two heavy hadrons $H_1$ and $H_2$, which reads
\be \label{eq:lifetimeratio}
{\tau(H_1)\over \tau(H_2)}= 1+{\Gamma_2-\Gamma_1\over \Gamma_1} &=& 1+ {\mu_\pi^2(H_1)-\mu^2_\pi(H_2) \over 2m_Q^2}+{c_{3,Q}+2c_{5,Q}\over c_{3,Q}}{\mu_G^2(H_1)-\mu^2_G(H_2) \over 2m_Q^2}   \non \\
&+& {c_{6,Q}\over c_{3,Q}} {\la H_2|T_6|H_2\ra\over 2m^3_Q m_{_{H_2}}}-{c_{6,Q}\over c_{3,Q}} {\la H_1|T_6|H_1\ra\over 2m^3_Q m_{_{H_1}}}  \non \\
&+& {c_{7,Q}\over c_{3,Q}} {\la H_2|T_7|H_2\ra\over 2m^4_Q m_{_{H_2}}}-{c_{7,Q}\over c_{3,Q}} {\la H_1|T_7|H_1\ra\over 2m^4_Q m_{_{H_1}}}+\cdots,
\en
where use of Eqs. (\ref{eq:NLrate}) and (\ref{eq:SLrate}) has been made,
$c_{3,Q}\equiv c_{3,Q}^{\rm NL}+c_{3,Q}^{\rm SL}$ and likewise for $c_{5,Q}$.
Note that the lifetime ratio computed in this manner is valid for $B$ mesons and bottom baryons where the HQE in $1/m_b$ converges nicely, but not for charmed hadrons where the inclusive rates are not dominated by the $c_{3,c}$ term.

\section{Lifetimes of heavy mesons}

\subsection{Lifetimes of bottom mesons}

We shall first fix the $b$ quark mass from the measured
inclusive semileptonic decay rate. Experimentally \cite{PDG},
\be \label{eq:Bsemi}
&& \B(B^+\to X_c e^+\nu_e)=(10.8\pm0.4)\%, \qquad \B(B^0\to X_c e^+\nu_e)=(10.1\pm0.4)\%, \non \\
&& \B(B^+/B^0\,{\rm admixture}\to X_ce^+\nu_e)=(10.65\pm0.16)\%.
\en
Theoretically, Eq. (\ref{eq:SLrate}) leads to
\footnote{Corrections to inclusive semileptonic $B$ decays have been calculated to order $\O(\alpha_s^2)$ and $\O(1/m_b^5)$ \cite{Mannel:2010}.}
\be \label{eq:BSLrate}
\Gamma(B\to X_ce^+\nu_e)
&=& \,{G_F^2m_b^5\over 192\pi^3}|V_{cb}|^2
\Bigg\{ I_0(x,0,0) \left(1+a{\alpha_s\over \pi}\right)\left(1-{\mu^2_\pi\over 2m_b^2}\right) \non \\
&+& \left(I_0(x,0,0)-4I_1(x,0,0)\right)\left(1+b{\alpha_s\over \pi}\right){\mu^2_G\over 2m_b^2}\Bigg\},
\en
where we have included the radiative corrections to order $\alpha_s$ characterized by the parameters $a$ and $b$. The order $\alpha_s$ corrections alone without $\mu^2_\pi/m_b^2$ or $\mu^2_G/m_b^2$  terms were first calculated in \cite{Hokim,Nir}. Corrections of order $\alpha_s\mu^2_\pi/m_b^2$ have been calculated in \cite{Becher,Alberti:mupi}, while the $\O(\alpha_s\mu^2_G/m_b^2)$ terms  in \cite{Alberti:muG,Mannel:muG}.
The analytic expression of the coefficient $a$ can be found in \cite{Nir} and the $b$ term in \cite{Mannel:muG}.

The inclusive rate is very sensitive to the quark mass $m_b$. The reliability of the calculation depends on the ability to control the higher order
contributions in the double series expansion in $\alpha_s$ and $\Lambda_{\rm QCD}/m_b$. The pole mass definition for heavy quark masses does not converge very well and moreover it is plagued by the renormalon ambiguity \cite{Bigi:renormalon,Beneke:renormalon}. For the short-distance $\overline{\rm MS}$ mass $\bar m_b(\bar m_b)$, it is not under good control for the smaller scale $\mu\sim 1$ GeV. Two different schemes commonly used to define the short-distance $b$-quark mass are the kinetic \cite{Bigi:kin,Bigi:1997} and the 1S \cite{Hoang} schemes. We follow \cite{Gambino:2016} for a recent global fit of inclusive semileptonic $B$ decays in the kinetic scheme. This analysis includes higher power corrections $\O(1/m_b^4)$ and $\O(m_b^5)$ and next-to-leading order QCD (NLO-QCD) corrections $\O(\alpha_s^2)$. In this scheme, it is conventional to constrain the charm quark mass to be the $\overline{\rm MS}$ one $\bar m_c({\rm 3\,GeV})=0.987\pm0.013$ GeV which yields a better convergence of the perturbative series. The results of the fit are \cite{Gambino:2016}:
\be \label{eq:masskin}
m_b^{\rm kin}({\rm 1 GeV})=4.546\pm0.021~{\rm GeV},\quad \mu^2_\pi=0.432\pm0.068~{\rm GeV}^2, \quad \mu^2_G=0.355\pm0.060~{\rm GeV}^2. \non \\
\en
The corresponding $\overline{\rm MS}$ mass $\bar m_b(\bar m_b)$  for $m_b^{\rm kin}({\rm 1 \,GeV})$ is close to the usual one.

To the leading-order QCD (LO-QCD), the definition of the quark mass is very arbitrary. If everything is calculated consistently to NLO-QCD, the dependence of the final result on the quark mass definition will be considerably weak when the relations between different quark mass schemes are used consistently at the NLO accuracy.
\footnote{Besides the above-mentioned inclusive semileptonic $B$ decays, another example is given in \cite{Krinner} for inclusive nonleptonic decay rates to NLO which are calculated in various quark mass schemes. The numerical results are similar for different short-distance quark masses.}
Although dimension 3, 4 and 6 Wilson coefficients up to NLO-QCD are available for heavy $B$ and $D$ mesons, they are still absent for heavy baryons. Dimension-7 Wilson coefficients are known only to the LO level for both heavy mesons and baryons. For this reason, in this work we shall focus on the LO-QCD study. In the bottom hadron sector, we use the quark masses $m_b=4.546$ GeV and $m_c=0.987$ GeV obtained in Eq. (\ref{eq:masskin}).
The reason is that the calculated inclusive semileptonic rate to LO, $\Gamma(B\to X_c e^+\nu_e)=4.59\times 10^{-14}$ GeV using the kinetic $b$ quark mass is very close to  the experimental measurement:
\be \label{eq:SLexpt}
\Gamma(B^+/B^0\,{\rm admixture}\to X_ce^+\nu_e) &=& (4.476\pm 0.067)\times 10^{-14}
{\rm GeV},
\en
where the average lifetime $\tau(B^+/B^0/B_s/b{\rm-baryon}\,{\rm admixture})=(1.566\pm0.003)$ ps \cite{PDG} and branching fraction  (\ref{eq:Bsemi})  have been made.  If the running quark masses $\bar m_b(\bar m_b)=4.248$ GeV and $\bar m_c(\bar m_c)=1.277$ GeV  are employed, the obtained $\Gamma(B\to X_ce^+\nu_e)$ to LO will be too small by 47\%  compared to experiment. It should be stressed that to the NLO-QCD, the dependence of the inclusive semileptonic rate on the quark mass definition is considerably weak.

We next turn to the spectator effects of order $1/m_b^3$.
The $W$-exchange contributions to $B_d$ and $B_s$ correspond to the Pauli interference terms $\T_{6,int}^{\B_Q,q_2}$ and $\T_{6,int}^{\B_Q,q_3}$, respectively, in Eq. (\ref{eq:T6baryon}) for heavy baryon decays, while
the Pauli interference in inclusive nonleptonic $B^-$ decay corresponds to the annihilation term $\T_{6,ann}^{\B_Q,q_1}$ in heavy baryon decays:
\be  \label{eq:T6meson}
\T_{6,ann}^{B_d} &=& -{G_F^2m_b^2\over 6\pi}\,\xi (1-x)^2\Bigg\{
 \left({1\over N_c}c_1^2+2c_1c_2+N_cc_2^2\right)\Big[ (1+{x\over 2})(\bar bd)(\bar db)\non \\
&-& (1+2x)\bar b(1-\gamma_5)d\bar d(1+\gamma_5)b\Big]
 +2c_1^2\Big[ (1+
{x\over 2})(\bar bt^ab)(\bar dt^ad)\non \\
&-& (1+2x)\bar b(1-\gamma_5)t^ad
\bar d(1+\gamma_5)t^ab\Big]\Bigg\},    \non \\
\T_{6,ann}^{B_s} &=& -{G_F^2m_b^2\over 6\pi}\,\xi \sqrt{1-4x}\Bigg\{
 \left({1\over N_c}c_1^2+2c_1c_2+N_cc_2^2\right)\Big[ (1-x)(\bar bs)(\bar sb)\non \\
&-& (1+2x)\bar b(1-\gamma_5)s\bar s(1+\gamma_5)b\Big]
 +2c_1^2\Big[ (1-x)(\bar bt^ab)(\bar st^as)\non \\
&-& (1+2x)\bar b(1-\gamma_5)t^as
\bar s(1+\gamma_5)t^ab\Big]\Bigg\},    \\
\T_{6,int}^{B_u} &=& {G^2_Fm_b^2\over 6\pi}\,\xi\,(1-x)^2\Bigg\{ \left(2N_c c_1c_2+
c_1^2+c_2^2\right)(\bar b u)(\bar ub)\non \\
&+& 2N_c(c_1^2+c_2^2)(\bar bt^au)(\bar u t^ab)\Bigg\}, \non
\en
where $(\bar q_1 t^a q_2)\equiv \bar q_1\gamma_\mu(1-\gamma_5)t^aq_2$ with $t^a=\lambda^a/2$ and
we have applied the relation
\be \label{eq:ta}
(\bar Q_\alpha t^a_{\alpha\beta}q_\beta)(\bar q_\rho t^a_{\rho\sigma}Q_\sigma)={1\over 2}(\bar QQ)(\bar qq)-{1\over 2N_c}(\bar Qq)(\bar qQ)
\en
to the transition operators so that they are more suitable for the matrix element evaluation in the meson case. Likewise, dimension-7 operators relevant for heavy meson decays can be read from Eq. (\ref{eq:T7baryon_original}):
\be  \label{eq:T7meson}
\T_{7,ann}^{B_d} &=& {G_F^2m_b^2\over 6\pi}\,\xi (1-x)\Bigg\{  \left({1\over N_c}c_1^2+2c_1c_2+N_cc_2^2\right)\Big[ -(1-x)(1+2x)(P_1^d+P_2^d) \non \\
&+& 2(1+x+x^2)P_3^d-12x^2P_4^d -(1-x)(1+{x\over 2})P_5^d+(1-x)(1+2x)P_6^d\Big] \non \\
&+& 2c_1^2 \Big[ -(1-x)(1+2x)(S_1^d+S_2^d)+2(1+x+x^2)S_3^d-12x^2S_4^d  \non \\
&-& (1-x)(1+{x\over 2})S_5^d+(1-x)(1+2x)S_6^d\Big]
 \Bigg\},    \\
\T_{7,ann}^{B_s} &=& {G_F^2m_Q^2\over 6\pi}\,\xi \sqrt{1-4x}\Bigg\{  \left({1\over N_c}c_1^2+2c_1c_2+N_cc_2^2\right)\Big[ -(1+2x)(P_1^s+P_2^s) \non \\
&+&{2\over 1-4x}(1-2x-2x^2)P_3^s - {24x^2\over 1-4x}P_4^s -(1-x)P_5^{s}+(1+2x)P_6^{s} \Big] \non \\
&+& 2c_1^2 \Big[ -(1+2x)(S_1^s+S_2^s)+{2\over 1-4x}(1-2x-2x^2)S_3^s
-{24x^2\over 1-4x}S_4^s  \non \\
&-& (1-x)S_5^{s}+(1+2x)S_6^{s} \Big] \Bigg\}, \non \\
\T_{7,int}^{B_u} &=& {G^2_Fm_b^2\over 6\pi}\,\xi\,(1-x)\Bigg\{ \left(2N_c c_1c_2+c_1^2+c_2^2\right)\Big[2(1+x)P_3^u+(1-x)P^u_5\Big] \non \\
&+& 6(c_1^2+c_2^2)\Big[2(1+x)S_3^u+(1-x)S^u_5\Big]  \Bigg\}.   \non
\en

For the meson matrix elements of four-quark operators, we follow
\cite{Neubert97} to define the bag parameters $B_i$ and $\ep_i$ to parametrize the hadronic matrix elements in a model-independent way:
\be \label{eq:m.e.T6meson}
&& \la B_q|(\bar bq)(\bar qb)| B_q\ra = \,f_{B_q}^2m_{B_q}^2 B_1,
\non \\
&& \la  B_q|\bar b(1-\gamma_5)q\bar  q(1+\gamma_5)b| B_q\ra =\,f_{
B_q}^2m_{B_q}^2 B_2, \non \\
&& \la  B_q|(\bar b\,t^aq)(\bar q\,t^ab)|
 B_q\ra = \,f_{B_q}^2m_{B_q}^2 \ep_1,  \\
&& \la  B_q| b\,t^a(1-\gamma_5)q\bar q\,t^a
(1+\gamma_5)b| B_q\ra = \,f_{B_q}^2m_{B_q}^2 \ep_2.   \non
\en
Under the vacuum-insertion approximation,
bag parameters are given by $B_i=1$ and
$\ep_i=0$, but they will be treated as free parameters here. In the large-$N_c$ limit, it is expected that $B_i\sim{\cal O}(1)$ and $\ep_i\sim {\cal O}(1/N_c)$.
Likewise,
the matrix elements of dimension-7 four-quark operators read \cite{Lenz:D}
\be \label{eq:m.e.T7meson}
\la  B_q|P_i^q| B_q\ra &=& -{m_q\over m_b}f_B^2m_B^2\rho_i^q, \qquad i=1,2, \non \\
\la  B_q|P_i^q| B_q\ra &=& (-1)^if_B^2m_B^2{1\over 2}\left({m_B^2\over m_b^2}-1\right)\rho_i^q, \qquad i=3,4,  \\
\la  B_q|P_i^q| B_q\ra &=& (-1)^if_B^2m_B^2{1\over 2}\left({m_B^2\over m_b^2}-1\right)\rho_i^q, \qquad i=5,6,  \non
\en
and similar parametrization for the color-octet operators with the replacement of $P\to S$ and $\rho_i\to\sigma_i$. Under the vacuum-insertion approximation, $\rho_i^q=1$ and all $\sigma$'s vanish.

Applying Eqs. (\ref{eq:m.e.T6meson}) and (\ref{eq:m.e.T7meson}) to evaluate the $B$-meson matrix elements of dimension-6 and dimension-7 four-quark operators, (\ref{eq:T6meson}) and (\ref{eq:T7meson}), respectively,  the spectator effects
\be
\Gamma^{\rm spec}(B_q)={\la B_q|\T_6+\T_7|B_q\ra\over 2m_{B_q}}
\en
have the expressions
\be \label{eq:Bspect}
\Gamma^{\rm ann}(B_d) &=& -{G_F^2m_b^2\over \pi}|V_{cb}V_{ud}|^2|\psi^B_{b\bar q}(0)|^2 (1-x)^2 \Bigg\{ \left({1\over N_c}c_1^2+2c_1c_2+N_cc_2^2\right)\Bigg[ (1+{x \over 2})B_1  \non \\
&-&(1+2x)B_2
+ \left( {1+x+x^2\over 1-x}\rho_3^d+{6x^2\over 1-x}\rho_4^d-{1\over 2}(1+{x\over 2})\rho_5^d-{1\over 2}(1+2x)\rho_6^d\right) \non \\
&\times& \left({m_B^2\over m_b^2}-1\right)\Bigg]
+ 2c_1^2 \Bigg[ (1+{x \over 2})\ep_1  -(1+2x)\ep_2\
 +\Bigg( {1+x+x^2\over 1-x}\sigma_3^d  \non \\
&+& {6x^2\over 1-x}\sigma_4^d
 - {1\over 2}(1+{x\over 2})\sigma_5^d-{1\over 2}(1+2x)\sigma_6^d
 \Bigg)\left({m_B^2\over m_b^2}-1\right)\Bigg] \Bigg\}, \non \\
\Gamma^{\rm ann}(B_s) &=& -{G_F^2m_b^2\over \pi}|V_{cb}V_{cs}|^2|\psi^{B_s}_{b\bar s}(0)|^2 \sqrt{1-4x} \Bigg\{ \left({1\over N_c}c_1^2+2c_1c_2+N_cc_2^2\right)\Bigg[ (1-x)B_1   \\
&-&(1+2x)B_2
+ \left( {1-2x-2x^2\over 1-4x}\rho_3^s+{12x^2\over 1-4x}\rho_4^s -{1-x\over 2}\rho^s_5-{1+2x\over 2}\rho^s_6\right)  \non \\
&\times& \left({m_{B_s}^2\over m_b^2}-1\right)\Bigg]
+ 2c_1^2 \Bigg[ (1-x)\ep_1  -(1+2x)\ep_2
 +\Bigg( {1-2x-2x^2\over 1-4x}\sigma_3^s  \non \\
&+& {12x^2\over 1-4x}\sigma_4^s-{1-x\over 2}\sigma^s_5-{1+2x\over 2}\sigma^s_6\Bigg)
\left({m_{B_s}^2\over m_b^2}-1\right)\Bigg]  \Bigg\}, \non \\
\Gamma^{\rm int}(B_u) &=& {G_F^2m_b^2\over \pi}|V_{cb}V_{ud}|^2|\psi^B_{b\bar q}(0)|^2(1-x)^2\Bigg\{ \left(2N_c c_1c_2+c_1^2+c_2^2\right)\Bigg[ B_1-\left({1+x\over 1-x}\rho_3^u+{1\over 2}\rho_5^u\right) \non \\
&\times& \left({m_B^2\over m_b^2}-1\right)\Bigg]
+ 6(c_1^2+c_2^2)\Bigg[\ep_1
 -\left({1+x\over 1-x}\sigma_3^u+{1\over 2}\sigma_5^u\right) \left({m_B^2\over m_b^2}-1\right)\Bigg]\Bigg\}, \non
\en
where $|\psi^B_{b\bar q}(0)|^2={1\over 12}f_B^2m_B$ is the $B$ meson wave function at the origin squared. Since $(m^2_B/m^2_b-1)\sim \O(1/m_b)$, it is evident that contributions from dimension-7 operators are suppressed by $\Lambda/m_b$ relative to the dimension-6 ones.
As stressed in \cite{Neubert97},
the coefficients of $B_i$ in $\Gamma^{\rm ann}(B_d)$ are one to two orders of magnitude
smaller than that of $\ep_i$. Therefore, contributions of $B_i$ can be
safely neglected at least in $\Gamma^{\rm ann}(B_d)$.
There exist several estimates of the bag parameters $B_i$ and $\ep_i$ based on sum rules \cite{Chernyak,Colangelo,Baek,Cheng-Yang} and lattice QCD \cite{DiPierro,Becirevic}.
On the basis of HQET sum rules, 
\footnote{The HQET sum rule calculation in the literature relies
on the work of \cite{Grozin:2008nu} where the necessary three-loop HQET
diagrams have been computed and on the work of \cite{Grozin:2016uqy} where these results have been first used for an estimate of the
bag parameter.}
these parameters have been updated recently to be \cite{Kirk}
\be \label{eq:bagHQET}
B_1=1.028^{+0.064}_{-0.056}, \quad B_2=0.988^{+0.087}_{-0.079}, \quad \ep_1=-0.107^{+0.028}_{-0.029}, \quad \ep_2=-0.033^{+0.021}_{-0.021},
\en
evaluated at the $\mu=\bar m_b(\bar m_b)$ scale. For the parameters $\rho_i^q$ and $\sigma_i^q$ we shall use the vacuum-insertion estimates, namely $\rho_i^q=1$ and $\sigma_i^q=0$.

To compute the nonleptonic decay rate we apply the Wilson coefficient
functions
\be
c_1(\mu)=1.14,~~~~~c_2(\mu)=-0.31,
\en
which are evaluated at $\mu=4.4$ GeV to the leading logarithmic approximation
(see Table XIII of \cite{Buras}). The total rate reads
\be
\Gamma=\Gamma^{\rm dec}+\Gamma^{\rm ann}+\Gamma^{\rm int}+\Gamma^{\rm semi},
\en
where the decay rate of the heavy quark $b$ of the $B$ meson is given by
\be \label{eq:dec}
\Gamma^{\rm dec}(B) &=& {G_F^2m_b^5\over 192\pi^3}\,\xi\,
\Bigg\{ c_{3,b}^{\rm NL} \Big[1-{\lambda_1\over 2m_b^2}+{d_B\lambda_2\over 2m_b^2}\Big]+ 2c_{5,b}^{\rm NL}{d_B\lambda_2\over m_b^2}\Bigg\}
\en
with $d_B=3$.
Now we make a comparison with \cite{Lenz:2014} on the $b$ quark lifetime. From Eqs. (\ref{eq:c3,5b}) and (\ref{eq:c3,5bSL}) we obtain $c_{3,b}=c_{3,b}^{\rm NL}+c_{3,b}^{\rm SL}=5.61$ to LO-QCD level for $m_b=4.546$ GeV and $m_c=0.987$ GeV, while Lenz got $c_{3,b}=5.29\pm0.35$ to LO-QCD for $\bar m_b(\bar m_b)=4.248$ GeV, $\bar m_c(\bar m_b)=0.997$ GeV and $6.88\pm0.74$ with NLO-QCD corrections. We have the lifetime of a free $b$ quark $\tau_b=1.51$ ps, while Lenz obtained $\tau_b=(1.65\pm0.24)$ ps \cite{Lenz:2014}, where
\be
\Gamma_b = {G_F^2m_b^5\over 192\pi^3}|V_{cb}|^2 c_{3,b}.
\en
The calculated lifetimes of $B$ mesons shown in Table \ref{tab:Blifetime} are longer than the free $b$ quark lifetime for two reasons: (i) $1/m_b^2$ effects characterized by $\lambda_1$ and $\lambda_2$ will suppress the nonleptonic rate slightly, and (ii)  inclusive semileptonic rate is slightly suppressed by QCD corrections, the $a$ and $b$ terms in Eq. (\ref{eq:BSLrate}).

\begin{table}
\caption{Various contributions to the decay rates (in units of
$10^{-13}$ GeV) of $B$ mesons.  Experimental values are taken from \cite{PDG}.}
\label{tab:Blifetime}
\begin{center}
\begin{tabular}{|c c c c  l l  l  l|} \hline \hline
 & $\Gamma^{\rm dec}$ & $\Gamma^{\rm ann}$ & $\Gamma^{\rm int}_-$ & ~ $\Gamma^{\rm semi}$ & ~$\Gamma^{\rm tot}$ &
~$\tau(10^{-12}s)$~ & ~ $\tau_{\rm expt}(10^{-12}s)$~ \\
\hline
 $B^+$ & 3.102 & 0 & $-0.267$ & ~1.000 &
~3.834~ & ~ 1.717~ & ~$1.638\pm0.004$~  \\
 $B^0_d$~ & ~3.102~ & ~0.039~ & ~$0$~ &  ~1.000~ &
~4.141~ & ~  1.590~ &  ~$1.520\pm0.004$~   \\
 $B_s^0$~ & ~3.102~ & ~0.053~ & ~$0$~ &  ~1.000~ &
~4.155~ & ~ 1.584 ~ &  ~$1.510\pm0.005$~   \\
\hline \hline
\end{tabular}
\end{center}
\end{table}

Eq. (\ref{eq:Bspect}) implies a constructive $W$-exchange
to $B_d$ and $B_s$ and a destructive Pauli interference to $B_u$. From Eq. (\ref{eq:lifetimeratio}) we obtain model-independent expressions for the lifetime ratios
\be \label{eq:timerat:anal}
{\tau(B^+)\over \tau(B^0_d)} &=& 1+(0.037B_1+0.0008B_2-0.57\epsilon_1+0.15\epsilon_2)_{\rm dim-6}
+(-0.015\rho_3-0.0064\rho_5  \non \\
&& ~ -0.00014\rho_6  +0.11\sigma_3-0.0007\sigma_4+0.099\sigma_5+0.026\sigma_6)_{\dim-7}, \\
{\tau(B^0_s)\over \tau(B^0_d)} &=& 1+(0.0003B_1-0.0005B_2+0.060\epsilon_1-0.079\epsilon_2)_{\rm dim-6}+(0.0002\rho_3-0.0001\rho_5 \non \\
&& ~ -0.0001\rho_6 +0.039\sigma_3+0.002\sigma_4-0.015\sigma_5-0.019\sigma_6)_{\rm dim-7}, \non
\en
where we have decomposed the lifetime ratios in terms of dimension-6 and dimension-7 contributions and dropped the superscripts of $\rho_i^q$ and $\sigma_i^q$ by assuming their flavor independence for simplicity.
Using Eq. (\ref{eq:bagHQET}) for dimension-6 bag parameters and the  vacuum-insertion approximation for dimension-7 $\rho_i^q$ and $\sigma_i^q$, 
\footnote{We have followed \cite{Kirk} to assign fixed uncertainties to both $\rho_i$ and $\sigma_i$, namely $\rho_i=1\pm1/2$ and $\sigma_i=0\pm1/6$.}
we obtain
\be \label{eq:Bratio}
&&\left.{\tau(B^+)\over \tau(B^0_d)}\right|_{\rm theo}=1.074^{+0.017}_{-0.016},   \qquad\quad~~ \left.{\tau(B^+)\over \tau(B^0_d)}\right|_{\rm expt}=1.076\pm0.004, \non \\
&&\left.{\tau(B^0_s)\over \tau(B_d^0)}\right|_{\rm theo}=0.9964\pm0.0024,    \qquad \left.{\tau(B^0_s)\over \tau(B_d^0)}\right|_{\rm expt}=0.994\pm0.004,
\en
to be compared with $\left.{\tau(B^+)\over \tau(B_d)}\right|_{\rm theo}=1.082^{+0.022}_{-0.026}$ and  $\left.{\tau(B_s)\over \tau(B_d)}\right|_{\rm theo}=0.9994\pm0.0025$ found in \cite{Kirk}. 
The theoretical uncertainties in (\ref{eq:Bratio}) arise mainly from the bag parameters given in Eq. (\ref{eq:bagHQET}).
Our results are in excellent agreement with experiment. If we apply naive vacuum-insertion approximation also to dimension-6 bag parameters, we will have
\be
\left.{\tau(B^+)\over \tau(B^0_d)}\right|_{\rm VIA}=1.016, \qquad \left.{\tau(B^0_s)\over \tau(B_d^0)}\right|_{\rm VIA}=1.000\,.
\en
This implies that the main contribution to the $B_u-B_d$ lifetime ratio arises from the color-octet terms $-0.57\epsilon_1+0.15\epsilon_2+\cdots$ in Eq. (\ref{eq:timerat:anal}). The predicted $\epsilon_1$ and $\epsilon_2$ from HQET sum rules given in Eq. (\ref{eq:bagHQET}) yield an excellent description of lifetime ratios. Note that $\epsilon_1\approx 3\epsilon_2$ here rather than $\epsilon_1\approx 0.3\epsilon_2$ as originally argued in \cite{Neubert97}.

Several remarks are in order. (i) Weak annihilation contributions to $B_d$ and $B_s$ are suppressed relative to the Pauli interference due to a large cancelation between the bag parameters $B_1$ and $B_2$ and the partial cancelation between $\epsilon_1$ and $\epsilon_2$, see Eq. (\ref{eq:Bspect}) and Table \ref{tab:Blifetime}. (ii) The annihilation contribution to $B_s$ is larger than that of $B_d$ owing to SU(3) breaking in the decay constants and masses. This explains why the lifetime $B_s$ is slightly shorter than $B_d$. (iii) To order $1/m_c^3$, we obtain $\tau(B^+)/\tau(B^0_d)=1.0945$, which can be checked from Eq. (\ref{eq:timerat:anal}). Hence, it is necessary to introduce dimension-7 operators in order to improve the agreement with experiment.

\subsection{Lifetimes of charmed mesons}
The  semileptonic inclusive decay $D\to X_s e^+\nu_e$, the analog of $B\to X_c e^+\nu_e$, has not been measured. Instead, what we have are \cite{PDG}
\be
&& \B(D^+\to Xe^+\nu_e)=(16.07\pm0.30)\%, \qquad \B(D^0\to Xe^+\nu_e)=(6.49\pm0.11)\%, \non \\
&&  \B(D^0_s\to Xe^+\nu_e)=(6.5\pm0.4)\%.
\en
We begin with the inclusive semileptonic decay rate of the $D$ meson given by Eq. (\ref{eq:SLrate})
\be \label{eq:Dsl}
\Gamma(D\to Xe^+\nu_e) &=& {G_F^2m_c^5\over 192\pi^3}\Bigg(|V_{cs}|^2 \eta(x,0,0)\left\{ c_{3,c}^{\rm SL}(x,0)\Big[1+{\lambda_1+d_D\lambda_2\over 2m_c^2}\Big]+ 2c_{5,c}^{\rm SL}(x,0){d_D\lambda_2\over m_c^2}\right\} \non \\
&+& |V_{cd}|^2 \eta(0,0,0)\left\{ c_{3,c}^{\rm SL}(0,0)
\Big[1+{\lambda_1+d_D\lambda_2\over 2m_c^2}\Big]+ 2c_{5,c}^{\rm SL}(0,0){d_D\lambda_2\over m_c^2}\right\} \Bigg),
\en
where $x=(m_s/m_c)^2$, $d_D=3$ and the Wilson coefficients $c_{3,c}^{\rm SL}$ and $c_{5,c}^{\rm SL}$ are given by Eq. (\ref{eq:c3,5cSL}). In the above equation we have included the radiative correction $\eta(x,0,0)$ given by the $(1+a\alpha_s/\pi)$ term in Eq. (\ref{eq:BSLrate}).
We find that the
experimental values for $D^+$ and $D^0$ semileptonic widths can be
fitted by the charm quark mass $m_c=1.56$ GeV.
\footnote{The semileptonic widths of $D^+$ and $D^0$ are very similar, while the $D_s^+$ one is smaller by 15\%.}
For the Wilson coefficients, we shall use the lowest order values
\be \label{eq:charmWc}
c_1(\mu)=\,1.346,~~~~~c_2(\mu)=-0.636
\en
evaluated at the scale $\mu=1.25$ GeV with $\Lambda^{(4)}_{\overline {\rm MS}}=325$ MeV (see Tables VI and VII of \cite{Buras}).

Just as the $B$ meson case, the spectator effects in the $D$ meson sector read
\be \label{eq:Dspect}
\Gamma^{\rm ann}(D^0) &=& -{G_F^2m_c^2\over 12\pi}|V_{cs}V_{ud}|^2 f_D^2m_D (1-x)^2 \Bigg\{ \left({1\over N_c}c_1^2+2c_1c_2+N_cc_2^2\right)\bigg[ (1+{x \over 2})B_1  \non \\
&-&(1+2x)B_2
+ \left( {1+x+x^2\over 1-x}\rho_3^u+{6x^2\over 1-x}\rho_4^u-{1\over 2}(1+{x\over 2})\rho_5^u-{1\over 2}(1+2x)\rho_6^u\right) \non \\
&\times& \left({m_D^2\over m_c^2}-1\right)\bigg]
+ 2c_1^2 \bigg[ (1+{x \over 2})\ep_1  -(1+2x)\ep_2\
 +\Bigg( {1+x+x^2\over 1-x}\sigma_3^u  \non \\
&+& {6x^2\over 1-x}\sigma_4^u
- {1\over 2}(1+{x\over 2})\sigma_5^u-{1\over 2}(1+2x)\sigma_6^u
 \Bigg)\left({m_D^2\over m_c^2}-1\right)\bigg] \Bigg\}, \non  \\
\Gamma^{\rm int}(D^+) &=& {G_F^2m_c^2\over 12\pi}|V_{cs}V_{ud}|^2 f_D^2m_D (1-x)^2\Bigg\{ \left(2N_c c_1c_2+c_1^2+c_2^2\right)\Bigg[ B_1-\left({1+x\over 1-x}\rho_3^d+{1\over 2}\rho_5^d\right) \non \\
&\times& \left({m_D^2\over m_c^2}-1\right)\Bigg]
+ 6(c_1^2+c_2^2)\Bigg[\ep_1
 -\left({1+x\over 1-x}\sigma_3^d+{1\over 2}\sigma_5^d\right) \left({m_D^2\over m_c^2}-1\right)\Bigg]\Bigg\},  \\
\Gamma^{\rm ann}(D_s^+) &=& -{G_F^2m_c^2\over 12\pi}|V_{cs}|^2 f_{D_s}^2m_{D_s}  \Bigg\{ \left({1\over N_c}c_2^2+2c_1c_2+N_cc_1^2)\right)|V_{ud}|^2\Bigg[ B_1 - B_2 -{1\over 2}(\rho^s_1+\rho^s_2){m_s\over m_c} \non \\
&+& \left(\rho^s_3-{1\over 2}(\rho^s_5+\rho^s_6)\right)
\left({m_{D_s}^2\over m_c^2}-1\right)\Bigg]+ 2c_2^2 |V_{ud}|^2\Bigg[ \ep_1  -\ep_2 -{1\over 2}(\sigma^s_1+\sigma^s_2){m_s\over m_c} \non \\
&+& \Big( \sigma^s_3
- {1\over 2}(\sigma^s_5+\sigma^s_6)
 \Big) \left({m_{D_s}^2\over m_c^2}-1\right)\Bigg]
+\Big[1+(1-z)^2(1+{z\over 2})\Big]B_1 \non \\
&-& \Big[1+(1-z)^2(1+2z)\Big]B_2
 - \Big[1+(1-z)^2(1+2z)\Big](\rho^s_1+\rho^s_2){m_s\over m_c} \non \\
&+& \left( [1+(1-z)(1+z+z^2)]\rho^s_3+6z^2(1-z)\rho^s_4\right)
\left({m_{D_s}^2\over m_c^2}-1\right) \Bigg\}, \non
\en
where $z=m_\mu^2/m_c^2$. We have followed \cite{Lenz:D} to derive the expression for the inclusive rate of $D_s^+$. Note that the contributions involving the $z$ terms arise from the leptonic intermediate states.

It is well known that $D^+$ has a longer lifetime than $D^0$ because of destructive Pauli interference \cite{Guberina79,Bilic}. To a good approximation to $1/m_c^3$ expansion, we have
\be
\Gamma(D^+)&\approx&  {G_F^2m_c^5\over 192\pi^3}\left[3(c_1^2+c_2^2)+2c_1c_2\right]+{G_F^2m_c^2\over 12\pi}(c_1^2+c_2^2+6c_1c_2)f_D^2m_D+\Gamma^{\rm semi}, \non \\
\Gamma(D^0)&\approx&  {G_F^2m_c^5\over 192\pi^3}\left[3(c_1^2+c_2^2)+2c_1c_2\right]+\Gamma^{\rm semi},
\en
where $\Gamma^{\rm semi}\approx G_F^2m_c^5/(96\pi^3)$.  For the decay constant $f_D$ of order 205 MeV (see \cite{FLAG} for a review), it is easily seen that the Pauli interference $\Gamma^{\rm int}(D^+)$ to order $1/m_c^3$ overcomes the $c$ quark decay rate so that $\Gamma(D^+)$ becomes negative no matter which charmed quark mass is employed, the $\overline{\rm MS}$ mass $\bar m_c(\bar m_c)=1.279$ GeV or the fit mass $m_c=1.56$ GeV. This remains to be true even if other sets of the bag parameters are used so long as they are not far from the vacuum insertion expectation.

In the literature, the lifetime ratio is often computed using the relation
\footnote{It is not meaningful to apply Eq. (\ref{eq:lifetimeratio}) to compute the lifetime ratio $R_D$  because (i) the $\Gamma(D^+)$ rate is not dominated by the $c_3$ term, and (ii) when Eq. (\ref{eq:lifetimeratio}) is applied to the ratio $R'_D$, it will lead to $R'_D=2-R_D$ which is negative.}
\be \label{eq:Dratio}
R_D\equiv {\tau(D^+)\over \tau(D^0)}=1+[\Gamma(D^0)-\Gamma(D^+)]\tau(D^+)_{\rm expt},
\en
where the experimental value of $\tau(D^+)$ is utilized on the r.h.s. of the above equation. However, it is important to keep in mind that the calculated $D^+$ lifetime is negative to order $1/m_c^3$. Hence, it does not make sense to apply the HQE to ${\cal O}(1/m_c^3)$ to predict a ``positive" lifetime ratio $\tau(D^+)/\tau(D^0)$
in spite of a negative $D^+$ lifetime predicted by the HQE at this level. Therefore, we should also consider the ratio
\be \label{eq:Dratio1}
R'_D\equiv {\tau(D^0)\over \tau(D^+)}=1+[\Gamma(D^+)-\Gamma(D^0)]\tau(D^0)_{\rm expt}
\en
to ensure that $R'_D=1/R_D$.
Their experimental values are given by \cite{PDG}
\be
\left.{\tau(D^+)\over\tau(D^0)}\right|_{\rm expt}=2.536\pm0.019, \qquad \left.{\tau(D^0)\over\tau(D^+)}\right|_{\rm expt}=0.394\pm0.003\,.
\en

It follows from Eqs. (\ref{eq:Dspect}), (\ref{eq:Dratio}) and (\ref{eq:Dratio1}) that
\be \label{eq:RD}
R_D &=& 1+(2.88B_1+0.11B_2-17.25\epsilon_1+3.71\epsilon_2)_{\rm dim-6} \non \\
&&~ \, +(-1.34\rho_3-0.62\rho_5+0.024\rho_6+4.25\sigma_3+3.70\sigma_5+0.80\sigma_6)_{\rm dim-7},  \\
R'_D &=& 1+(-1.13B_1-0.043B_2 +6.80\epsilon_1-1.46\epsilon_2)_{\rm dim-6} \non \\
&&~\, +(0.53\rho_3+0.24\rho_5-0.0093\rho_6  -1.68\sigma_3+1.46\sigma_5-0.31\sigma_6)_{\rm dim-7}.  \non
\en
We can use the experimental values of $R_D$ and $R'_D$ to constrain the bag parameters.
To order $1/m_c^3$,  we find $R_D=3.98$ and $R'_D=-0.18$ for $B_i=1$ and $\epsilon_i=0$. This implies a negative $D^+$ lifetime.
Indeed, the calculated $\tau(D^+)$ is $-8.4\times 10^{-13}s$.
The subleading $1/m_c$ corrections to the Pauli interference term, namely, $\Gamma^{\rm int}_7$ obtained from dimension-7 four-quark operators at $1/m_c^4$ level, contributes constructively to the $D^+$ width (see Eq. (\ref{eq:Dspect})).
Hence, it is conceivable that the  $1/m_c$ corrections to the Pauli interference will be able to render $\Gamma(D^+)$ positive in certain rages of the bag parameters.
With $m_c=1.56$ GeV, we find $R_D=2.06$ and $R'_D=0.58$ in the presence of $1/m_c^4$ corrections with $\rho_i=1$ and $\sigma_i=0$.

For the $D_s^+$ meson, we follow \cite{Lenz:D} to define a substracted $D_s^+$ lifetime by
\be
\bar\tau(D_s^+)={\tau(D_s^+)\over 1-\B(D_s^+\to\tau^+\nu_\tau)}=(0.533\pm0.004)\times 10^{-12}s
\en
as the decay $D_s^+\to\tau^+\nu_\tau$ cannot be properly described by HQE due to its small energy release. In analogue to the $D^+$ meson, we consider the lifetime ratios
\be \label{eq:Dsratio}
R_{D_s}&\equiv&  {\bar\tau(D^+_s)\over \tau(D^0)}=1+[\Gamma(D^0)-\bar\Gamma(D^+_s)]\bar \tau(D^+_s)_{\rm expt}, \non \\
R'_{D_s}&\equiv&  {\tau(D^0)\over \bar \tau(D^+_s)}=1+[\bar\Gamma(D^+_s)-\Gamma(D^0)]\tau(D^0)_{\rm expt},
\en
with the experimental values \cite{PDG}
\be
R_{D_s}=1.30\pm0.01 , \qquad R'_{D_s}=0.77\pm0.01\,.
\en
We obtain
\be \label{eq:RDs}
R_{D_s} &=& 1+(4.81B_1-4.82B_2-1.21\epsilon_1+1.22\epsilon_2)_{\rm dim-6}
 +(-0.17\rho_1-0.17\rho_2+2.86\rho_3   \non \\
&& ~\, -0.94\rho_5-0.94\rho_6
 -0.017\sigma_1-0.017\sigma_2-0.41\sigma_3+0.21\sigma_5+0.21\sigma_6)_{\rm dim-7}, \non \\
R'_{D_s} &=& 1+(-3.70B_1+3.71B_2 +0.93\epsilon_1-0.94\epsilon_2)_{\rm dim-6}
 +(0.13\rho_1+0.13\rho_2-2.20\rho_3 \non \\
 && ~\, +0.72\rho_5+0.72\rho_6 +0.013\sigma_1+0.013\sigma_2 +0.32\sigma_3-0.16\sigma_5-0.16\sigma_6)_{\rm dim-7}.
\en


In principle, if the vacuum-insertion expectation for $\rho$'s and $\sigma$'s is assumed, the four unknown parameters $B_{1,2}$ and $\epsilon_{1,2}$ can be obtained by solving the four equations for $R_D$, $R'_D$, $R_{D_s}$  and $R'_{D_s}$ given in Eqs. (\ref{eq:RD}) and (\ref{eq:RDs}). In practice, the solutions are very sensitive to the experimental values within errors. We pick up the solutions not far from the vacuum-insertion expectation.
\footnote{Some solutions, for example, $B_1=0.138, B_2=0.144,  \epsilon_1=-0.296$ and $\epsilon_2=-0.552$ can also reproduce the data, but they are ruled out as $B_{1,2}$ are too small, whereas $\epsilon_{1,2}$ are too large.}
For example, one of the solutions is
\be \label{eq:bagOur}
B_1=0.840, \quad B_2=0.919, \quad \epsilon_1=-0.060, \quad \epsilon_2=-0.025\,,
\en
at the scale $\mu=m_c$.
They reproduce the experimental values of $R_D$, $R'_D$, $R_{D_s}$ and $R'_{D_s}$. Note that in order to accommodate the experimental values  of $R_{D_s}$ and $R'_{D_s}$, it is necessary to have $B_2>B_1$. To see this, we take vacuum-insertion values for $\rho$'s and $\sigma$'s and find from Eq. (\ref{eq:RDs}) that
\be
R_{D_s} &=& 1.65+(4.81B_1-4.82B_2-1.21\epsilon_1+1.22\epsilon_2)_{\rm dim-6},
 \non \\
R'_{D_s} &=& 0.50+(-3.70B_1+3.71B_2 +0.93\epsilon_1-0.94\epsilon_2)_{\rm dim-6}.
\en
It is clear that if $B_2\approx B_1$ and $\epsilon_2\approx \epsilon_1$, the value of $R_{Ds}$ ($R'_{D_s}$) will be too large (small) compared to the data.
Hence, one needs $B_2>B_1$ and $|\epsilon_1|>|\epsilon_2|$ in order to suppress $R_{D_s}$ and enhance $R'_{D_s}$ simultaneously.

Our results are to be compared with the recent estimates based on HQET sum rules by Kirk, Lenz and Rauh (KLR) \cite{Kirk}
\be \label{eq:DbagLenz}
B_1=0.902^{+0.077}_{-0.051}, \quad B_2=0.739^{+0.124}_{-0.073}, \quad \epsilon_1=-0.132^{+0.041}_{-0.046}, \quad \epsilon_2=-0.005^{+0.032}_{-0.032},
\en
evaluated at the scale $\mu=3$ GeV. While KLR have updated the prediction of $\tau(D^+)/\tau(D^0)$, they did not perform the similar update for the lifetime ratio of $D_s^+$ and $D^0$ mesons. As discussed above, the explanation of the $R_{D_s}$ data requires that $B_1<B_2$.

Finally, we notice that the size of the subleading $1/m_c$ corrections
is $[(\Gamma(D^0)-\Gamma(D^+)]_{\rm dim-7}/[(\Gamma(D^0)-\Gamma(D^+)]_{\rm dim-6}\approx -56\%$, which is compatible with a convergent series.

\section{Lifetimes of heavy baryons}

\subsection{Lifetimes of bottom baryons}
The spectator effects in inclusive heavy bottom baryon decays arising from dimension-6 and dimension-7 operators are given by Eqs. (\ref{eq:T6baryon}) and (\ref{eq:T7baryon}), respectively.
We shall rely on the quark model to evaluate the baryon matrix elements of four-quark operators. In \cite{Cheng:1997} we have studied the matrix elements in the MIT bag model \cite{MIT,MIT_1,MIT_2,MIT_3} and the nonrelativistic quark model (NQM).
In analogue to Eq. (\ref{eq:m.e.T6meson}), we parameterize the four baryon matrix elements in a model-independent manner:
\footnote{This is similar to the hadronic parameters defined in Eq. (28) of \cite{Beneke:2002}.}
\be
&& \la \B_b|(\bar bq)(\bar qb)|\B_b\ra = f_{{B_q}}^2m_{_{B_q}}m_{_{\B_b}} L_1^{\B_b},  \non \\
&& \la \B_b|\bar b(1-\gamma_5)q\bar q(1+\gamma_5)b|\B_b\ra = f_{{B_q}}^2m_{_{B_q}}m_{_{\B_b}} L_2^{\B_b},  \\
&& \la \B_b|(\bar bb)(\bar qq)|\B_b\ra = f_{{B_q}}^2m_{_{B_q}}m_{_{\B_b}} L_3^{\B_b}, \non \\
&& \la \B_b|\bar b^\alpha(1-\gamma_5)q^\beta\bar q^\beta(1+\gamma_5)b^\alpha|
\B_b\ra = f_{{B_q}}^2m_{_{B_q}}m_{_{\B_b}} L_4^{\B_b}, \non
\en
where $\B_b$ stands for the antitriplet bottom baryon $T_b$ ($\Lambda_b$ or $\Xi_b$) or the sextet bottom baryon $\Omega_b$, $f_{B_q}$ and $m_{_{B_q}}$ are the decay constant and the mass of the heavy meson $B_q$, respectively. The four hadronic parameters $L_1,\cdots,L_4$ are not all independent.

First, since the color wavefunction for a baryon is totally antisymmetric,
the matrix element of $(\bar bb)(\bar qq)$ is the same as that of $(\bar bq)
(\bar qb)$ except for a sign difference. Thus we follow \cite{Neubert97} to
define a parameter $\tilde B$
\be \label{eq:Btilde}
L_3^{\B_b}=-\tilde{B}L_1^{\B_b},  \qquad L_4^{\B_b}=-\tilde{B}L_2^{\B_b}, \qquad {\rm for}~\B_b=T_b,\Omega_b,
\en
so that $\tilde B=1$ in the valence-quark approximation. Second, in the quark model evaluation we obtain (see Appendix B for derivation) \cite{Cheng:1997}
\be \label{eq:m.e.baryon}
\la T_b|(\bar bq)(\bar qb)|T_b\ra/(2m_{T_b}) &=& \cases{-\left|\psi^{T_b}_{bq}(0)\right|^2, & NQM \cr -(a_q+b_q), & MIT} \non \\
\la \Omega_b|(\bar bs)(\bar sb)|\Omega_b\ra/(2m_{\Omega_b}) &=& \cases{ -6\left| \psi^{\Omega_b}_{bs}(0)\right|^2, & NQM \cr -{1\over 3}(18a_s+2b_s+32c_s), & MIT}
\en
and
\be \label{eq:m.e.baryon_1}
\la T_b|\bar b(1-\gamma_5)q\bar q(1+\gamma_5)b|T_b\ra /(2m_{T_b})
&=& \cases{ {1\over 2}\left|\psi^{T_b}_{bq}(0)\right|^2, & NQM \cr   {1\over 2}(a_q+b_q), & MIT}  \non \\
\la \Omega_b|\bar b(1-\gamma_5)s\bar s(1+
\gamma_5)b|\Omega_b\ra /(2m_{\Omega_b})
&=& \cases{ -\left|\psi^{\Omega_b}_{bs}(0)\right|^2, & NQM   \cr
-(a_s-{5\over 3}b_s-{16\over 3}c_s), & MIT}
\en
where $a_q$, $b_q$ and $c_q$ are the four-quark overlap integrals used in the MIT bag model:
\be \label{eq:overlap}
a_q &=& \int d^3r\left[\,u_q^2(r)u_b^2(r)+v_q^2(r)v_b^2(r)\right],   \non \\
b_q &=& \int d^3r\left[\,u_q^2(r)v_b^2(r)+v_q^2(r)u_b^2(r)\right],   \non \\
c_q &=& \int d^3r\,u_q(r)v_q(r)u_b(r)v_b(r),
\en
which are expressed in terms of the large and small components $u(r)$ and
$v(r)$, respectively, of the quark wavefunction
(see e.g., Ref.~\cite{CT} for the technical detail of the bag model
evaluation). In deriving Eq. (\ref{eq:m.e.baryon_1}), use of
\be \label{eq:m.e.1}
\la T_b|\bar b^\alpha\gamma_\mu\gamma_5 b^\beta\bar q^\beta\gamma^\mu(1-
\gamma_5)q^\alpha|T_b\ra/(2m_{T_b}) &=& 0,   \non  \\
\la \Omega_b|\bar b^\alpha\gamma_\mu\gamma_5 b^\beta\bar s^\beta\gamma^\mu(1-
\gamma_5)s^\alpha|\Omega_b\ra/(2m_{\Omega_b}) &=& \cases{ 4\left| \psi^{\Omega_b}_{bs}(0)\right|^2 & NQM \cr 4\left(a_s-{b_s\over 3}\right) & MIT}
\en
and
\be \label{eq:rel}
\bar b^\alpha\gamma_\mu\gamma_5 b^\beta\bar q^\beta\gamma^\mu(1-\gamma_5)
q^\alpha=-\bar b(1-\gamma_5)q\bar q(1+\gamma_5)b-{1\over 2}(\bar bq)(\bar
q b),
\en
has been made. The first relation in Eq. (\ref{eq:m.e.1}) is a model-independent consequence of heavy quark spin symmetry \cite{Neubert97}. It follows from Eqs. (\ref{eq:Btilde}),  (\ref{eq:m.e.baryon}) and (\ref{eq:m.e.baryon_1}) that
\be
L_2^{T_b}=-{1\over 2}L_1^{T_b}, \qquad L_2^{\Omega_b}={1\over 6}L_1^{\Omega_b},
\en
where the second relation for the $\Omega_b$ is exact in the NQM but only an approximation in the MIT bag model.

Since the small component $v(r)$ is negligible in the NQM, baryon matrix elements of four-quark operators in the NQM and MIT models are the same  except for the replacement:
\be \label{eq:replacement}
a_q\to \int d^3r\,u_b^2(r)u_q^2(r),\qquad b_q\to 0,\qquad c_q\to 0.
\en
Hence, in the NQM $a_q$ is nothing but the baryon wave function at the origin squared $|\psi_{bq}(0)|^2$.
In general, the strength of destructive Pauli interference and $W$-exchange
is governed by $a_q+b_q$ in the bag model and $|\psi(0)|^2$ in the NQM.
However,  the bag model calculation
of $a_q+b_q$ generally gives a much smaller value than the  nonrelativistic estimate
of $|\psi(0)|^2$. As argued in \cite{Cheng:1997}, the difference between $a_q+b_q$ and $|\psi(0)|^2$ is not simply attributed
to relativistic corrections; it arises essentially from the
distinction in the spatial scale of the wavefunction especially at the origin.
As a consequence, both models give a quite different quantitative
description for processes sensitive to $|\psi(0)|^2$. It turns out that the NQM works better for heavy baryon decays. Hence, we will follow \cite{Cheng:1997} to consider the NQM estimate of baryon
matrix elements.

To estimate the bottom baryon wave function in the center, consider $|\psi^{\Lambda_b}_{bq}(0)|^2$ as an example.
A calculation of hyperfine splittings between $\Sigma_b$ and $\Lambda_b$ as well as between $B^*$ and $B$ based on the mass formula given in \cite{Rujula} yields \cite{Cortes:1980jz}
\be
|\psi^{\Lambda_b}_{bq}(0)|^2=\,{2m_q\over m_b-m_q}\,{m_{\Sigma_b}-
m_{\Lambda_b}\over m_{B^*}-m_B}\,|\psi^B_{b\bar q}(0)|^2,
\en
where the equality $|\psi^{\Sigma_b}_{bq}(0)|^2=|\psi^{\Lambda_b}_{bq}(0)|^2$
has been assumed. As a consequence, the wave function of a bottom baryon at the origin can be related to that of a $B$ meson.
Another approach proposed by Rosner \cite{Rosner} is to consider the hyperfine splittings of $\Sigma_b$ and $B$ separately so that
\be \label{eq:Psibq}
|\psi^{\Lambda_b}_{bq}(0)|^2=|\psi^{\Sigma_b}_{bq}(0)|^2=\,{4\over 3}\,{m
_{\Sigma^*_b}-m_{\Sigma_b}\over m_{B^*}-m_B}\,|\psi^B_{b\bar q}(0)|^2.
\en
This method is presumably more reliable as $|\psi_{bq}(0)|^2$ thus
determined does not depend on $m_b$ and the constituent quark mass $m_q$ directly.
Defining the wave function ratio
\be \label{eq:rb}
r_{\Lambda_b}=\left|{\psi^{\Lambda_b}_{bq}(0)\over \psi^B_{b\bar q}(0)}\right|^2,
\en
and noting that $|\psi^B_{b\bar q}(0)|^2={1\over 12}f_B^2m_B$, we see from Eqs. (\ref{eq:m.e.baryon}), (\ref{eq:m.e.baryon_1}) and (\ref{eq:m.e.1}) that
the parameters $L_1,\cdots,L_4$ for bottom baryons now read
\be \label{eq:m.e.dim6}
&& L_1^{T_b}=-{1\over 6}r_{_{T_b}}, \quad L_2^{T_b}={1\over 12}r_{_{T_b}}, \quad L_3^{T_b}={1\over 6}\tilde{B}r_{_{T_b}}, \quad L_4^{T_b}=-{1\over 12}\tilde{B}r_{_{T_b}}, \non \\
&& L_1^{\Omega_b}=-r_{_{\Omega_b}}, \quad ~L_2^{\Omega_b}=-{1\over 6}r_{_{\Omega_b}}, \quad L_3^{\Omega_b}=\tilde{B}r_{_{\Omega_b}}, \quad ~ L_4^{\Omega_b}={1\over 6}\tilde{B}r_{_{\Omega_b}}.
\en
Hence, the baryon matrix elements are expressed in terms of two independent parameters $r_{_{\B_c}}$ and $\tilde B$.

For dimension-7 four-quark operators, the baryon matrix elements are given by
\be \label{eq:m.e.dim7}
&& \la T_b|P_1^q|T_b\ra=\la T_b|P_2^q|T_b\ra={1\over 48} f_{B_q}^2 m_{_{B_q}}m_{_{T_b}}r_{_{T_b}}\left( {m^2_{T_b}-m^2_{diq}\over m_b^2}-1\right)\eta^q_{1,2},  \non \\
&& \la T_b|P_3^q|T_b\ra=-2\la T_b|P_4^q|T_b\ra=-{1\over 24} f_{B_q}^2 m_{_{B_q}}m_{_{T_b}}r_{_{T_b}}\left( {m^2_{T_b}-m^2_{diq}\over m_b^2}-1\right)\eta^q_{3,4}, \\
&& \la \Omega_b|P_1^s|\Omega_b\ra=\la \Omega_b|P_2^s|\Omega_b\ra={1\over 8} f_{B_s}^2 m_{_{B_s}}m_{_{\Omega_b}}r_{_{\Omega_b}}\left( {m^2_{\Omega_b}-m_{\{ss\}}^2\over m_b^2}-1\right)\eta^s_{1,2},  \non \\
&& \la \Omega_b|P_3^s|\Omega_b\ra=6\la \Omega_b|P_4^s|\Omega_b\ra=-{1\over 4} f_{B_s}^2 m_{_{B_s}}m_{_{\Omega_b}}r_{_{\Omega_b}}\left( {m^2_{\Omega_b}-m^2_{\{ss\}}\over m_b^2}-1\right)\eta^s_{3,4}, \non
\en
where $m_{diq}$ is the mass of the scalar diquark of $T_b$ and the parameters $\eta^q_i$ are expected to be of order unity.
We shall  follow \cite{Ebert} to use $m_{[ud]}=710$ MeV for $\Lambda_b$, $m_{[us]}=948$ MeV for $\Xi_b^0$, $m_{[ds]}=948$ MeV for $\Xi_b^-$  and $m_{\{ss\}}=1203$ MeV for $\Omega_b$ with $[qq']$ antisymmetric in flavor and $\{ss\}$ symmetric in flavor denoting scalar and axial-vector diquarks, respectively. For the matrix elements of the operators $\tilde P_i^q$, we follow Eq. (\ref{eq:Btilde}) to introduce a parameter $\tilde\beta_i^q$
\be \label{eq:beta}
\la \B_b|\tilde P_i^q|\B_b\ra=-\tilde \beta_i^q \la \B_b|P_i^q|\B_b\ra,
\en
so that $\tilde\beta_i^q=1$ under the valence quark approximation.

Two remarks are in order. First, unlike the meson matrix elements $\la B_q| P_{1,2}^q| B_q\ra$ in Eq. (\ref{eq:m.e.T7meson}) which are explicitly of order $m_q/m_b$ because of the definition of the operators $P_{1,2}^q$, the baryon matrix elements $\la \B_b|P_{1,2}^q|B_b\ra$ in Eq. (\ref{eq:m.e.dim7}) are not explicitly proportional to $m_q/m_b$. Nevertheless, it is easily seen that  $(m^2_{T_b}-m^2_{diq})/m_b^2-1$, for example, is indeed of order $m_q/m_b$. Second, unlike the meson case we do not know how to evaluate the baryon matrix elements of $P^q_{5,6}$. Since the operators $P^q_{5,6}$ arise from by expressing the QCD four-quark operators in terms of HQET operators
\be
\bar b\Gamma q\bar q\Gamma b=\bar h_v\Gamma q\bar q\Gamma h_v+{1\over 2m_b}\left[ \bar h_v(-i\stackrel{\leftarrow}{D\!\!\!\!/})\Gamma q\bar q\Gamma h_v+\bar h_v\Gamma q\bar q\Gamma(iD\!\!\!\!/)h_v\right],
\en
we shall follow \cite{Gabbiani:2003pq,Gabbiani:2004tp} to assume that dimension-7 operators contain full QCD $b$ quark fields. Therefore, to evaluate the baryon matrix elements of dimension-7 operators given in Eq. (\ref{eq:T7baryon}), we will drop the operators $P^q_{5,6}$ and $\tilde P_{5,6}^q$.

    To estimate the hadronic parameter $r_{_{\B_Q}}$ in the NQM,
we find from Eq. (\ref{eq:Psibq}) that
\be
r_{\Lambda_b}=\,{4\over 3}\,{m_{\Sigma_b^*}-m_{\Sigma_b}\over m_{B^*}-m_B},\qquad
r_{\Xi_b}=\,{4\over 3}\,{m_{\Xi_b^*}-m_{\Xi'_b}\over m_{B^*}-m_B},\qquad
r_{\Omega_b}=\,{4\over 3}\,{m_{\Omega_b^*}-m_{\Omega_b}\over m_{B^*}-m_B},
\en
and likewise for $r_{\Lambda_c},~r_{\Xi_c}$ and $r_{\Omega_c}$. Notice that the heavy-quark spin-violating mass relation \cite{Jenkins}
\be
(m_{\Sigma_Q^*}-m_{\Sigma_Q})+(m_{\Omega_Q^*}-m_{\Omega_Q})=2(m_{\Xi_Q^*}
-m_{\Xi'_Q})
\en
holds very accurately for $Q=b,c$. Numerically,  we obtain
\footnote{For a summary of the earlier estimates of $r_{_{\B_Q}}$, see \cite{Lenz:2014}.}
\be \label{eq:r}
&& r_{\Lambda_c}=0.610, \qquad
r_{\Xi_c}=0.656, \qquad
r_{\Omega_c}=0.664, \non \\
&& r_{\Lambda_b}=0.607, \qquad
r_{\Xi_b}=0.601, \qquad
r_{\Omega_b}=0.601,
\en
and
\be \label{eq:Psib}
&& |\psi^{\Lambda_b}_{bq}(0)|^2=0.92\times 10^{-2}{\rm GeV}^3,\qquad
|\psi^{\Xi_b}_{bq}(0)|^2=0.91\times 10^{-2}{\rm GeV}^3,   \non \\
&& |\psi^{\Omega_b}_{bs}(0)|^2=1.42\times 10^{-2}{\rm GeV}^3,
\en
for $f_{B_q}=186$ MeV and $f_{B_s}=230$ MeV \cite{FLAG}.
Therefore, the NQM estimate of $|\psi_{bq}^{\B_b}(0)|^2$ is
indeed larger than the analogous bag model quantity: $a_q+b_q\sim 3\times
10^{-3}{\rm GeV}^3$.

Except for the weak annihilation term, the expression of Pauli interference will be very lengthy if the hadronic parameters $\eta^q_i$ and $\tilde \beta^q_i$ are all treated to be different from each other. Since in realistic calculations we will set $\tilde \beta^q_i(\mu_h)=1$ under valence quark approximation and put $\eta^q_i$ to unity, we shall assume for simplicity that $\eta^q_i=\eta$ and $\tilde \beta^q_i=\tilde\beta$.
The spectator effects in nonleptonic decays of bottom baryons
are now readily obtained from Eqs. (\ref{eq:T6baryon}) and (\ref{eq:T7baryon}):
\be
\Gamma^{\rm ann}(\Lambda_b^0) &=& {G_F^2m_b^2\over 2\pi}\,|V_{cb}|^2\,
r_{{\Lambda_b}}\left|\psi^{B}_{b\bar q}(0)\right|^2 \Bigg\{\Big(\tilde B(c_1^2+c_2^2)-2c_1c_2\Big)(1-x)^2  \non \\
&+& {1\over 2}\Big(\tilde \beta(c_1^2+c_2^2)-2c_1c_2\Big)\eta(1-x^2)\left( {m^2_{\Lambda_b}-m^2_{[ud]}\over m_b^2}-1 \right)\Bigg\},   \non \\
\Gamma^{\rm ann}(\Xi_b^0) &=& {G_F^2m_b^2\over 2\pi}\,|V_{cb}|^2\,
r_{_{\Xi_b}}\left|\psi^{B}_{b\bar q}(0)\right|^2 \Bigg\{\Big(\tilde B(c_1^2+c_2^2)-2c_1c_2\Big)(1-x)^2  \non \\
&+& {1\over 2}\Big(\tilde \beta(c_1^2+c_2^2)-2c_1c_2\Big)\eta(1-x^2)\left( {m^2_{\Xi_b}-m^2_{[us]}\over m_b^2}-1 \right)\Bigg\},   \non \\
\Gamma^{\rm int}_-(\Lambda_b^0) &=& -{G_F^2m_b^2\over 4\pi}\,|V_{cb}V_{ud}|^2\, r_{_{\Lambda_b}}\left|\psi^{B}_{b\bar q}(0)\right|^2
\Bigg\{ \Big(\tilde Bc_1^2-2c_1c_2-N_cc_2^2\Big)\Bigg((1-x)^2(1+x)
\non \\
&+&  \left|{V_{cd}\over V_{ud}}\right|^2\sqrt{1-4x}\, \Bigg)
 -{1\over 2}\Big(\tilde \beta c_1^2-2c_1c_2-N_cc_2^2\Big)\eta(1-x)(1+x+2x^2)\left( {m^2_{\Lambda_b}-m^2_{[ud]}\over m_b^2}-1 \right)\Bigg\},   \non \\
\Gamma^{\rm int}_-(\Xi_b^0) &=& -{G_F^2m_b^2\over 4\pi}\,|V_{cb}V_{cs}|^2 \sqrt{1-4x}~r_{_{\Xi_b}}\left|\psi^{B}_{b\bar q}(0)\right|^2
\Bigg\{ \Big(\tilde Bc_1^2-2c_1c_2-N_cc_2^2\Big)  \non \\
&+& {1\over 6}\Big(\tilde \beta c_1^2-2c_1c_2-N_cc_2^2\Big)\eta\left(1+2x+{2(1-2x+10x^2)\over 1-4x}\right)\left( {m^2_{\Xi_b}-m^2_{[us]}\over m_b^2}-1 \right)\Bigg\},    \\
\Gamma^{\rm int}_-(\Xi_b^-) &=& -{G_F^2m_b^2\over 4\pi}\,|V_{cb}|^2\, r_{_{\Xi_b}}\left|\psi^{B}_{b\bar q}(0)\right|^2
\Bigg\{ \Big(\tilde Bc_1^2-2c_1c_2-N_cc_2^2\Big)\Big(|V_{ud}|^2(1-x)^2(1+x)
\non \\
&+& |V_{cs}|^2\sqrt{1-4x}\,\Big)
+\Big(\tilde \beta c_1^2-2c_1c_2-N_cc_2^2\Big)\eta\Bigg[-{1\over 2}|V_{ud}|^2(1-x)(1+x+2x^2) \non \\
&+& {1\over 6}|V_{cs}|^2\sqrt{1-4x}\left(1+2x+{2(1-2x+10x^2)\over 1-4x}\right)\Bigg]
\left( {m^2_{\Xi_b}-m^2_{[ds]}\over m_b^2}-1 \right)\Bigg\},   \non \\
\Gamma^{\rm int}_-(\Omega_b^-) &=& -{G_F^2m_b^2\over 6\pi}\,|V_{cb}V_{cs}|^2
\sqrt{1-4x}~r_{\Omega_b}\left|\psi^{B_s}_{b\bar s}(0)\right|^2\Bigg\{ \Big(\tilde{B}c_1^2-
2c_1c_2-N_cc_2^2\Big)(5-8x) \non \\
&+& {3\over 2}\Big(\tilde{\beta}c_1^2-
2c_1c_2-N_cc_2^2\Big)\eta\left(1+2x+ {2(1-2x-8x^2)\over 1-4x}\right)\left( {m^2_{\Omega_b}-m^2_{\{ss\}}\over m_b^2}-1 \right)\Bigg\},   \non
\en
where use has been made of Eqs. (\ref{eq:m.e.dim6}) and (\ref{eq:m.e.dim7}).
Note that there is no weak annihilation
contribution to the $\Xi_b^-$ and $\Omega_b$
and that there are two Cabibbo-allowed Pauli interference terms in $\Xi_b^-$
decay, and one Cabibbo-allowed as well as one Cabibbo-suppressed
interferences in $\Lambda_b$ decay.

To compute the decay widths of bottom baryons,
we have to specify the values of $\tilde B$ and $r$. Since $\tilde B=1$ in the
valence-quark approximation and since the wavefunction squared ratio $r$
is evaluated using the quark model, it is reasonable to assume that the NQM
and the valence-quark approximation are most reliable when the baryon matrix
elements are evaluated at a typical hadronic scale $\mu_{\rm had}$. As
shown in \cite{Neubert97}, the parameters $\tilde B$ and $r$ renormalized
at two different scales are related via the renormalization group equation
to be
\be \label{eq:RGE1}
\tilde B(\mu)r(\mu) =\, \tilde B(\mu_{\rm had})r(\mu_{\rm had}),  \qquad \tilde B(\mu) =\, {\tilde{B}(\mu_{\rm had})\over \kappa+{1\over N_c}(\kappa
-1)\tilde B(\mu_{\rm had}) }\,,
\en
with
\be \label{eq:RGE2}
\kappa=\left({\alpha_s(\mu_{\rm had})\over \alpha_s(\mu)}\right)^{3N_c/2
\beta_0}=\sqrt{\alpha_s(\mu_{\rm had})\over \alpha_s(\mu)}
\en
and $\beta_0={11\over 3}N_c-{2\over 3}n_f$. We consider the hadronic scale in the range of $\mu_{\rm had}\sim 0.65-1$ GeV. Taking  the medium scale $\mu_{\rm had}=0.825$ GeV as an illustration, we obtain $\alpha_s(\mu_{\rm had})
=0.59$, $\tilde{B}(\mu)=0.54\tilde B(\mu_{\rm
had})\simeq 0.54$ and $r(\mu)\simeq 1.86\,r(\mu_{\rm had})$.
The parameter $\tilde\beta$ is treated in a similar way.
Using the values of $r(\mu_{\rm had})$ given in Eq. (\ref{eq:r}), the calculated inclusive decay rates of bottom baryons are summarized in Table \ref{tab:bottombaryon}. We find that the lifetimes of bottom baryons stay almost constant with variation of $\mu_{\rm had}$.

\begin{table}[t]
\caption{Various contributions to the decay rates (in units of
$10^{-13}$ GeV) of bottom baryons with the hadronic scale $\mu_{\rm had}=0.825$ GeV.}
{
\begin{center} \label{tab:bottombaryon}
\begin{tabular}{|c c c c c c c c|} \hline
 & $\Gamma^{\rm dec}$ & $\Gamma^{\rm ann}$ & $\Gamma^{\rm int}_-$ &
$\Gamma^{\rm semi}$ & $\Gamma^{\rm tot}$ & $\tau(10^{-12}s)$~ & ~
$\tau_{\rm expt}(10^{-12}s)$~ \\  \hline
~$\Lambda_b^0$ & ~3.108~ & ~0.228~ & ~$-0.053$~ & ~1.055~ & ~4.338~ & ~1.517~
& ~$1.470\pm 0.010$~   \\
~$\Xi_b^0$ & 3.108 & 0.232 & $-0.084$ & 1.055 & 4.310 & 1.527 &  ~$1.479\pm 0.031$~    \\
~$\Xi_b^-$ & 3.108 & & $-0.130$ & 1.055 & 4.032 & 1.633 &  ~$1.571\pm 0.040$~   \\
~$\Omega_b^-$ & 3.105 & & $-0.341$ & 1.039 & 3.803 & 1.730 &   ~$1.64^{+0.18}_{-0.17}$~   \\
\hline
\end{tabular}
\end{center} }
\end{table}

We see from Table \ref{tab:bottombaryon} that the bottom baryon lifetimes follow the pattern
\be
\tau(\Omega_b^-)>\tau(\Xi_b^-)>\tau(\Xi_b^0)\simeq\tau(\Lambda_b^0).
\en
Theoretically, this pattern originates from the fact that while $\Lambda_b,~\Xi_b^0,~\Xi_b
^-,~\Omega_b$ all receive contributions from the destructive Pauli interference,
only $\Lambda_b$ and $\Xi_b^0$ have weak annihilation effects and that the destructive Pauli interference $\Gamma^{\rm int}_-$  in $\Omega_b$ is the largest due to the presence of two valence $s$ quarks in its quark content. The $\Xi_b^-$ has the second largest $\Gamma^{\rm int}_-$ due to the Pauli interference of identical $s$ quarks and the interference of identical $d$ quarks.

Several remarks are in order.
(i) There is a tiny difference between the semileptonic decays of the antitriplet $\Lambda_b$ or $\Xi_b$  and the sextet $\Omega_b$. It comes from the fact that the chromomagnetic
operator contributes to the matrix element of $\Omega_b$ but not to
$\Lambda_b$ or $\Xi_b$ as the light degrees of freedom in the latter are
spinless. (ii) It is evident from Tables \ref{tab:Blifetime} and \ref{tab:bottombaryon} that $W$-annihilation contribution in $B$ decays is much smaller than that in bottom baryon decays. The $W$-exchange in $B$ decays is helicity suppressed, while it is neither helicity nor color
suppressed in the heavy baryon case.  (iii) As pointed out in \cite{Beneke:2002}, $b$-flavor-conserving decays such as $\Xi_b^-\to\Lambda_b^0\pi^-,\Lambda_b^0 e^-\bar\nu_e$ and $\Xi_b^0\to\Lambda_b^0\pi^0$ could affect the total rates of the $\Xi_b$. These heavy-flavor-conserving weak decays were studied more than two decades ago within the framework that incorporates both heavy-quark and chiral symmetries \cite{ChengHFC,Cheng:2015}. The branching fraction of $\Xi_b\to\Lambda_b\pi$ is found to be of order $(0.1\sim 1)\%$, consistent with the recent LHCb measurement which lies in the range from $(0.57\pm0.21)\%$ to $(0.19\pm0.07)\%$ \cite{LHCb:HFC}. Hence, contributions from  $b$-flavor-conserving decays can be safely neglected for our present purpose.

From Eq. (\ref{eq:lifetimeratio}) we obtain the following lifetime ratios
\footnote{ The experimental lifetime ratio $\tau(\Xi_b^0)/\tau(\Xi_b^-)=0.929\pm0.028$ (or $\tau(\Xi_b^-)/\tau(\Xi_b^0)=1.076\pm0.032$) is quoted in the Heavy Flavor Averaging Group \cite{HFAG}.}
\be \label{eq:bbaryonratio}
&& \left.{\tau(\Xi_b^-)\over \tau(\Lambda_b^0)}\right|_{\rm theo}=1.073^{+0.009}_{-0.004},    \qquad \left.{\tau(\Xi_b^-)\over \tau(\Lambda_b^0)}\right|_{\rm expt}=1.089\pm0.028, \non \\
&& \left.{\tau(\Xi_b^-)\over \tau(\Xi_b^0)}\right|_{\rm theo} =1.066^{+0.009}_{-0.004},  \qquad  \left.{\tau(\Xi_b^-)\over \tau(\Xi_b^0)}\right|_{\rm expt}=1.083\pm0.036,  \\
&& \left.{\tau(\Omega_b^-)\over \tau(\Xi_b^-)}\right|_{\rm theo}=1.054^{+0.006}_{-0.002},   \qquad
\left.{\tau(\Omega_b^-)\over \tau(\Xi_b^-)}\right|_{\rm expt}=1.11\pm0.16, \non
\en
and
\be
\left.{\tau(\Lambda_b^0)\over \tau(B_d^0)}\right|_{\rm theo}=0.953^{+0.006}_{-0.008}, \qquad
\left.{\tau(\Lambda_b^0)\over \tau(B_d^0)}\right|_{\rm expt}=0.964\pm0.007\,.
\en
They are in good agreement with experiment \cite{PDG}.
The theoretical uncertainties for bottom baryon lifetime ratios can arise from many different places such as the nonperturbative parameters $\mu_\pi^2$ and $\mu_G^2$, where QCD sum rule and lattice calculations are still not available, and the matrix elements of dimension-6 and -7 operators. In the quark model, the unknown matrix elements are expressed in terms of two parameters $r_{\B_b}$, the wave function ratio, and $\tilde B$, which is equal to unity under the valence quark approximation. The estimate of the former is quite uncertain in the literature (see \cite{Lenz:2014} for a review). Therefore, it is far more difficult to estimate the uncertainties than the $B$ meson case. Nevertheless, there is one uncertainty which we can estimate reliably, namely, the hadronic scale $\mu_{\rm had}$ introduced before. The baryon matrix
elements need to be evaluated at a typical hadronic scale $\mu_{\rm had}$ in order to comply with the valence quark approximation.  We consider the hadronic scale in the range of $0.65-1$ GeV and use $\mu_{\rm had}=0.825\pm0.175$ GeV to estimate the uncertainties. The theoretical errors in Eq. (\ref{eq:bbaryonratio}) we have computed arise solely from the uncertainty of the hadronic scale. The uncertainty in the prediction of $\tau(\Lambda_c^+)/\tau(B_d^0)$ comes from the bag parameters in (\ref{eq:bagHQET}) and the hadronic scale $\mu_{\rm had}$. 
We see that the current world average of $\tau(\Lambda_b^0)/\tau(B_d^0)$ can be nicely explained within the framework of the HQE.

\subsection{Lifetimes of charmed baryons}
We first summarize the spectator effects relevant to charmed baryon decays derived from Eqs. (\ref{eq:T6baryon}) and (\ref{eq:T7baryon}):
\be \label{eq:Spectorcharmbary}
\Gamma^{\rm ann}(\Lambda_c^+) &=& {G_F^2m_c^2\over 2\pi}\,|V_{cs}V_{ud}|^2\,
r_{{\Lambda_c}}\left|\psi^{D}_{c\bar q}(0)\right|^2 \Bigg\{\Big(\tilde B(c_1^2+c_2^2)-2c_1c_2\Big)(1-x)^2  \non \\
&+& {1\over 2}\Big(\tilde \beta(c_1^2+c_2^2)-2c_1c_2\Big)\eta(1-x^2)\left( {m^2_{\Lambda_c}-m^2_{[ud]}\over m_c^2}-1 \right)\Bigg\},   \non \\
\Gamma^{\rm ann}(\Xi_c^+) &=& {G_F^2m_c^2\over 2\pi}\,|V_{cs}V_{us}|^2\,
r_{_{\Xi_c}}\left|\psi^{D}_{c\bar q}(0)\right|^2 \Bigg\{\Big(\tilde B(c_1^2+c_2^2)-2c_1c_2\Big)(1-x)^2  \non \\
&+& {1\over 2}\Big(\tilde \beta(c_1^2+c_2^2)-2c_1c_2\Big)\eta(1-x^2)\left( {m^2_{\Xi_c}-m^2_{[us]}\over m_c^2}-1 \right)\Bigg\},   \non \\
\Gamma^{\rm ann}(\Xi_c^0) &=& {G_F^2m_c^2\over 2\pi}\,|V_{cs}V_{ud}|^2\,
r_{_{\Xi_c}}\left|\psi^{D}_{c\bar q}(0)\right|^2 \Bigg\{\Big(\tilde B(c_1^2+c_2^2)-2c_1c_2\Big)(1-x)^2  \non \\
&+& {1\over 2}\Big(\tilde \beta(c_1^2+c_2^2)-2c_1c_2\Big)\eta(1-x^2)\left( {m^2_{\Xi_c}-m^2_{[ds]}\over m_c^2}-1 \right)\Bigg\},   \non \\
\Gamma^{\rm ann}(\Omega_c^0) &=& 3{G_F^2m_c^2\over \pi}\,|V_{cs}V_{us}|^2\,
r_{_{\Omega_c}}\left|\psi^{D_s}_{c\bar s}(0)\right|^2 \Bigg\{\Big(\tilde B(c_1^2+c_2^2)-2c_1c_2\Big)(1-x)^2  \non \\
&+& {1\over 2}\Big(\tilde \beta(c_1^2+c_2^2)-2c_1c_2\Big)\eta(1-x^2)\left( {m^2_{\Omega_c}-m^2_{\{ss\}}\over m_c^2}-1 \right)\Bigg\},   \non \\
\Gamma^{\rm int}_-(\Lambda_c^+) &=& -{G_F^2m_c^2\over 4\pi}\,|V_{cs}V_{ud}|^2\, r_{_{\Lambda_c}}\left|\psi^{D}_{c\bar q}(0)\right|^2
\Bigg\{ \Big(\tilde Bc_1^2-2c_1c_2-N_cc_2^2\Big)\Big((1-x)^2(1+x)
\non \\
&+&  \left|{V_{cd}\over V_{ud}}\right|^2\sqrt{1-4x}\, \Big)
 -{1\over 2}\Big(\tilde \beta c_1^2-2c_1c_2-N_cc_2^2\Big)\eta(1-x)(1+x+2x^2)\left( {m^2_{\Lambda_c}-m^2_{[ud]}\over m_c^2}-1 \right)\Bigg\},   \non \\
\Gamma^{\rm int}_-(\Xi_c^+) &=& -{G_F^2m_c^2\over 4\pi}\,|V_{cs}V_{ud}|^2 \sqrt{1-4x}~r_{_{\Xi_c}}\left|\psi^{D}_{c\bar q}(0)\right|^2
\Bigg\{ \Big(\tilde Bc_1^2-2c_1c_2-N_cc_2^2\Big)  \non \\
&+& {1\over 6}\Big(\tilde \beta c_1^2-2c_1c_2-N_cc_2^2\Big)\eta\left(1+2x+{2(1-2x+10x^2)\over 1-4x}\right)\left( {m^2_{\Xi_c}-m^2_{[us]}\over m_c^2}-1 \right)\Bigg\},    \\
\Gamma^{\rm int}_+(\Lambda_c^+) &=& {G_F^2m_c^2\over 4\pi}\,|V_{cd}V_{ud}|^2\, r_{_{\Lambda_c}}\left|\psi^{D}_{c\bar q}(0)\right|^2
\Bigg\{ \Big(2c_1c_2+N_cc_1^2-\tilde B c_2^2\Big)
\non \\
&-&  {1\over 2}\Big(2c_1c_2+N_cc_1^2-\tilde \beta c_2^2 \Big)\eta\left( {m^2_{\Lambda_c}-m^2_{[ud]}\over m_c^2}-1 \right)\Bigg\},   \non \\
\Gamma^{\rm int}_+(\Xi_c^+) &=& {G_F^2m_c^2\over 4\pi}\,|V_{cs}V_{ud}|^2\, r_{_{\Xi_c}}\left|\psi^{D}_{c\bar q}(0)\right|^2
\Bigg\{ \Big(2c_1c_2+N_cc_1^2-\tilde B c_2^2\Big)\Big(1+\left|{V_{us}\over V_{ud}}\right|^2(1-x)^2(1+x)\Big)
\non \\
&-&  {1\over 2}\Big(2c_1c_2+N_cc_1^2-\tilde \beta c_2^2\Big)\eta\Big(1+\left|{V_{us}\over V_{ud}}\right|^2(1-x)^2(1+x)\Big)\left( {m^2_{\Xi_c}-m^2_{[us]}\over m_c^2}-1 \right)\Bigg\},   \non \\
\Gamma^{\rm int}_+(\Xi_c^0) &=& {G_F^2m_c^2\over 4\pi}\,|V_{cs}V_{ud}|^2\, r_{_{\Xi_c}}\left|\psi^{D}_{c\bar q}(0)\right|^2
\Bigg\{ \Big(2c_1c_2+N_cc_1^2-\tilde B c_2^2\Big)\Big(1+\left|{V_{us}\over V_{ud}}\right|^2(1-x)^2(1+x)\Big)
\non \\
&-&  {1\over 2}\Big(2c_1c_2+N_cc_1^2-\tilde \beta c_2^2\Big)\eta\Big(1+\left|{V_{us}\over V_{ud}}\right|^2(1-x)^2(1+x)\Big)\left( {m^2_{\Xi_c}-m^2_{[ds]}\over m_c^2}-1 \right)\Bigg\},   \non \\
\Gamma^{\rm int}_+(\Omega_c^0) &=& {G_F^2m_c^2\over 6\pi}\,|V_{cs}V_{ud}|^2\, r_{_{\Omega_c}}\left|\psi^{D_s}_{c\bar s}(0)\right|^2
\Bigg\{ \Big(2c_1c_2+N_cc_1^2-\tilde B c_2^2\Big)\Big(5+\left|{V_{us}\over V_{ud}}\right|^2(1-x)^2(5+x)\Big)
\non \\
&-&  {9\over 2}\Big(2c_1c_2+N_cc_1^2-\tilde \beta c_2^2\Big)\eta\left( {m^2_{\Omega_c}-m^2_{\{ss\}}\over m_c^2}-1 \right)\Bigg\},   \non
\en
where $\tilde B$, $\eta$ and $\tilde\beta$  are the hadronic parameters defined in Eqs. (\ref{eq:Btilde}), (\ref{eq:m.e.dim7}) and (\ref{eq:beta}), respectively, and the wavefunction ratio $r_{\B_c}$ defined in  analog to Eq. (\ref{eq:rb}) with values given in Eq. (\ref{eq:r}). As stated before, we follow \cite{Ebert} to use $m_{[ud]}=710$ MeV, $m_{[us]}=m_{[ds]}=948$ MeV and $m_{\{ss\}}=1203$ MeV for the diquark masses.

Unlike bottom baryon decays, there exist constructive Pauli interference terms $\Gamma^{\rm int}_+$ in charmed baryon decays in addition to the destructive Pauli interference $\Gamma^{\rm int}_-$. Cabibbo-allowed (Cabibbo-suppressed) $\Gamma^{\rm int}_+$
arises from the constructive interference between the $s$ ($d$)
quark produced in the $c$ quark decay and the spectator $s$ ($d$) quark in the
charmed baryon (see Fig. \ref{fig:spectator}.(c)).

For the semileptonic inclusive decay of the charmed baryons meson,
the semileptonic decay rate has the same expression as Eq. (\ref{eq:Dsl})
except that the parameter $d_D$ is replaced by $d_{\B_c}$, which is equal to 0 for the antitriplet charmed baryons $\Lambda_c$, $\Xi_c$ and 4 for the $\Omega_c$.
For charmed baryons $\Xi_c$ and $\Omega_c$, there is an additional
contribution to the semileptonic width coming from the Pauli
interference of the $s$ quark \cite{Voloshin} (Fig. \ref{fig:CharmSL}). The dimension-6 contribution $\Gamma^{\rm SL}_{6,int}(\B_c)$ is given before by Eq. (\ref{eq:SL6}). As for the dimension-7 four-quark operator for semileptonic decays, it can be written as
\be
{\cal T}^{\rm SL}_7(\B_c)={G_F^2m_c^2\over 6\pi}|V_{cs}|^2\sum_{\ell=e,\mu}\left(\tilde g_i^{\nu \ell}P_i^s+\tilde h_i^{\nu\ell}S_i^s\right),
\en
with the coefficients given by \cite{Lenz:D}
\be
\tilde g_1^{\nu\mu}=-(1-z)^2(1+2z),  && \qquad \tilde g_2^{\nu\mu}=-(1-z)^2(1+2z), \non \\
\tilde g_3^{\nu\mu}=2(1-z)(1+z+z^2), && \qquad \tilde g_4^{\nu\mu}=-12z^2(1-z),
\en
and $\tilde h_i^{\nu\mu}=0$, where $z=(m_\mu/m_c)^2$. The coefficients $\tilde g_i^{\nu e}$ are given by setting $z=0$. Noting $\Gamma^{\rm SL}_{7,int}(\B_c)=\la \B_c|{\cal T}_7^{\rm SL}|\B_c\ra/(2m_{\B_c})$ and using Eqs. (\ref{eq:m.e.dim6}) and (\ref{eq:m.e.dim7}) to evaluate the baryon matrix elements, we obtain
\be
\Gamma^{\rm SL}_{int}(\Xi_c)&=& \Gamma^{\rm SL}_{6,int}(\Xi_c)+\Gamma^{\rm SL}_{7,int}(\Xi_c) \non \\
&=& {G_F^2m_c^2\over 4\pi}|V_{cs}|^2r_{_{\Xi_c}}|\psi^{D_s}_{c\bar s}(0)|^2\left[ 1-(1+{1\over 2}z^2-z^3)\left({m_{\Xi_c}^2-m_{[sq]}^2\over m_c^2}-1\right)\right],
\non \\
\Gamma^{\rm SL}_{int}(\Omega_c) &=& {G_F^2m_c^2\over 6\pi}|V_{cs}|^2r_{_{\Omega_c}}|\psi^{D_s}_{c\bar s}(0)|^2\left[ 5-9(1-{5\over 6}z^2+{1\over 3}z^3)\left({m_{\Omega_c}^2-m_{\{ss\}}^2\over m_c^2}-1\right)\right].
\en
We shall see later that, depending on the parameter $r$, the spectator
effect
in semileptonic decay of $\Xi_c$ and $\Omega_c$ can be very significant, in
particular for the latter.

We now turn to the heavy baryon wavefunction at the origin. We learn from Eq. (\ref{eq:Psib}) that $|\psi_{bq}^{\B_b}(0)|^2$ is of order $1\times 10^{-2}{\rm GeV}^3$. Likewise, $|\psi(0)|^2$ for hyperons is also of the same order of magnitude as the bottom baryons (see \cite{LaYaou} for details).  However, for the charmed baryon we obtain
$|\psi^{\Lambda_c}_{cq}(0)|^2=r_{\Lambda_c}|\psi^D_{c\bar q}(0)|^2=3.8\times 10^{-3}{\rm GeV}^3$  under the assumption that the
$D$ meson wavefunction in the center squared $|\psi^D_{c\bar q}(0)|^2$ is
identified with ${1\over 12}f_D^2m_D$. However, this is smaller than those in bottom or hyperon decays. This means that $|\psi^D_{c\bar q}(0)|^2$ is not simply equal to ${y\over 12}f_D^2m_D$ with $y=1$. We shall use $y=1.75$.

For the numerical results, we first consider the semileptonic decays. The measured inclusive semileptonic rate of the $\Lambda_c^+$
\be \label{eq:LamcSLexpt}
\Gamma(\Lambda_c^+\to Xe^+\nu_e)_{\rm expt}=\,(1.307\pm 0.112)\times
10^{-13}{\rm GeV},
\en
obtained from $\B(\Lambda_c^+\to Xe^+\nu_e)=(3.97\pm0.34)\%$, an average of the Mark II measurement of $(4.5\pm1.7)\%$ \cite{MarkII} and the recent BESIII result of $(3.95\pm0.35)\%$ \cite{BES:LamcSL}, is larger than that of $D$ mesons:
\be
\Gamma(D^+\to Xe^+\nu_e)_{\rm expt}=(1.017\pm0.019)\times 10^{-13}{\rm GeV}, \non \\
\Gamma(D^0\to Xe^+\nu_e)_{\rm expt}=(1.042\pm0.018)\times 10^{-13}{\rm GeV}.
\en
Theoretically, the difference between $\Lambda_c$ and $D$ comes from the $\lambda_2$ terms in Eq. (\ref{eq:Dsl}) which are absent in the former. Our prediction
\be \label{eq:LamcSL}
\Gamma(\Lambda_c^+\to Xe^+\nu_e)= 1.415\times 10^{-13}{\rm GeV},
\en
is consistent with experiment (\ref{eq:LamcSLexpt}).
Writing $\Gamma^{\rm SL}=\Gamma^{\rm SL}_c+\Gamma^{\rm SL}_{int}$, we see from Table \ref{tab:lifetime3_charmbary} that the spectator effects to ${\cal O}(1/m_c^3)$ in the semileptonic decays of $\Xi_c$ and $\Omega_c$ are quite significant, in particular for the latter.

\begin{table}[t]
\caption{Various contributions to the decay rates (in units of
$10^{-12}$ GeV) of singly charmed baryons to order $1/m_c^3$ with the hadronic scale $\mu_{\rm had}=0.825$ GeV.   Experimental values of charmed baryon lifetimes are taken from \cite{PDG}.}
\label{tab:lifetime3_charmbary}
\begin{center}
\begin{tabular}{|c c c c c l l  l  l|} \hline \hline
 & $\Gamma^{\rm dec}$ & $\Gamma^{\rm ann}$ & $\Gamma^{\rm int}_-$ &
$\Gamma^{\rm int}_+$ & ~ $\Gamma^{\rm semi}$ & ~$\Gamma^{\rm tot}$ &
~$\tau(10^{-13}s)$~ & ~ $\tau_{\rm expt}(10^{-13}s)$ \\
\hline
 ~$\Lambda_c^+$ & ~0.886~ & ~1.479~ & ~$-0.400$~ & 0.042 & ~0.215~ &
~2.221~ & ~  2.96~ &  ~$2.00\pm 0.06$~   \\
 ~$\Xi_c^+$ & 0.886 & 0.085 & $-0.431$ & 0.882 & ~0.726 &
~2.148~ & ~ 3.06~ & ~$4.42\pm0.26$  \\
 ~$\Xi_c^0$ & 0.886 & 1.591 & & 0.882 & ~0.726 &
~4.084 & ~ 1.61 & ~$1.12^{+0.13}_{-0.10}$ \\
 ~$\Omega_c^0$ & 1.019 & 0.515 & & 2.974 & ~1.901 &
~6.409 & ~ 1.03  & ~$0.69\pm 0.12$  \\
\hline \hline
\end{tabular}
\end{center}
\end{table}

   To proceed the hadronic decay rates, we employ the Wilson coefficients
given in Eq. (\ref{eq:charmWc}). As before,  we consider the hadronic scale in the range of $\mu_{\rm had}\sim 0.65-1$ GeV and obtain $\tilde{B}(\mu)=0.70\tilde B(\mu_{\rm
had})\simeq 0.70$ and $r(\mu)\simeq 1.42\,r(\mu_{\rm had})$ at the medium scale $\mu_{\rm had}=0.825$ GeV.
Repeating the same exercise as the bottom
baryon case, the results of calculations to order $1/m_c^3$ are exhibited in Table \ref{tab:lifetime3_charmbary}. Unlike the bottom baryon case where the lifetimes stay almost constant with variation of $\mu_{\rm had}$, the lifetimes of charmed baryons increase by around $10\%$ when the hadronic scale varies from 0.65 to 1.0 GeV. Nevertheless, the lifetime ratios remain nearly constant.

We see from Table \ref{tab:lifetime3_charmbary} the lifetime pattern
\be
\tau(\Xi_c^+)>\tau(\Lambda_c^+)>\tau(\Xi_c^0)>\tau(\Omega_c^0)
\en
is in accordance with experiment (for early studies of charmed baryon lifetimes, see \cite{Guberina:1986,SV:1986,BS93,Bellini,Melic,Guberina:2000,Guberina:2002,Guberina:2004}).
This lifetime hierarchy is  understandable qualitatively but not quantitatively. The $\Xi_c^+$ baryon is
longest-lived among charmed baryons because of the smallness of
$W$-exchange and partial cancellation between constructive and destructive
Pauli interferences, while $\Omega_c$ is shortest-lived due to the
presence of two $s$ quarks in the $\Omega_c$ that renders the contribution of
$\Gamma^{\rm int}_+$ largely enhanced. It is also clear from Table III that,
although the qualitative feature of the lifetime pattern is comprehensive, the
quantitative estimates of charmed baryon lifetimes and their ratios are
still rather poor. For example, $R_1\equiv\tau(\Xi_c^+)/\tau(\Lambda_c^+)$ and $R_2\equiv \tau(\Xi_c^+)/\tau(\Xi_c^0)$ are calculated to be 1.03 and 1.90, respectively, while experimentally $R_1=2.21\pm0.15$ and $R_2=3.95\pm0.47$.

\begin{table}[t]
\caption{Various contributions to the decay rates (in units of
$10^{-12}$ GeV) of singly charmed baryons after including subleading $1/m_c$ corrections to spectator effects.  The hadronic scale is chosen to be $\mu_{\rm had}=0.825$ GeV.
}
\label{tab:lifetime4_charmbary}
\begin{center}
\begin{tabular}{|c c c c r r l  l  l|} \hline \hline
 & $\Gamma^{\rm dec}$ & $\Gamma^{\rm ann}$ & $\Gamma^{\rm int}_-$ &
$\Gamma^{\rm int}_+$ & ~ $\Gamma^{\rm semi}$ & ~~$\Gamma^{\rm tot}$ &
~$\tau(10^{-13}s)$~ & ~ $\tau_{\rm expt}(10^{-13}s)$~ \\
\hline
 ~$\Lambda_c^+$ & ~0.886~ & ~2.179~ & ~$-0.211$~ & 0.022 & ~0.215 &
~~3.091~ & ~ 2.12~ &  ~$2.00\pm 0.06$~   \\
 ~$\Xi_c^+$ & 0.886 & 0.133 & $-0.186$ & 0.407 & ~0.437 &
~~1.677~ & ~ 3.92~ & ~$4.42\pm0.26$  \\
 ~$\Xi_c^0$ & 0.886 & 2.501 & $$ & 0.405 & ~0.435 &
~~4.228 & ~ 1.56 & ~$1.12^{+0.13}_{-0.10}$ \\
 ~$\Omega_c^0$ & 1.019 & 0.876 & & $-0.559$ & ~$-0.256$ &
~~1.079 & ~ 6.10  & ~$0.69\pm 0.12$  \\
\hline \hline
\end{tabular}
\end{center}
\end{table}

It is evident that, contrary to $B$ meson and bottom baryon cases where the HQE in $1/m_b$ leads to the lifetime ratios in excellent agreement with experiment, the heavy quark expansion in $1/m_c$ does not work well for describing the lifetime pattern of charmed baryons. Since the charm quark is not heavy enough, it is perhaps sensible to consider the subleading $1/m_c$ corrections to spectator effects as depicted in Eq. (\ref{eq:Spectorcharmbary}). The numerical results are shown in Table \ref{tab:lifetime4_charmbary}. By comparing Table \ref{tab:lifetime4_charmbary} with Table \ref{tab:lifetime3_charmbary}, we see that $\Gamma(\Lambda_c^+)$ is enhanced while $\Gamma(\Xi_c^+)$ is suppressed so that the resulting lifetime ratio $R_1$ is enhanced from 1.03 to 1.84. This means that $1/m_c$ corrections to spectator effects described by dimension-7 operators are in the right direction. However, the calculated $\Omega_c$ lifetime becomes entirely unexpected: the shortest-lived $\Omega_c$ turns out to be the longest-lived one to ${\cal O}(1/m_c^4)$. This is because the dimension-7 contributions $\Gamma^{\rm int}_{+,7}(\Omega_c)$ and $\Gamma^{\rm SL}_7(\Omega_c)$ are destructive and their size are so large that they overcome the dimension-6 ones and flip the sign. Of course, a negative $\Gamma^{\rm SL}(\Omega_c)$ does not make sense as the subleading corrections are too large to justify the validity of the HQE.

\begin{figure}[t]
\begin{center}
\includegraphics[height=47mm]{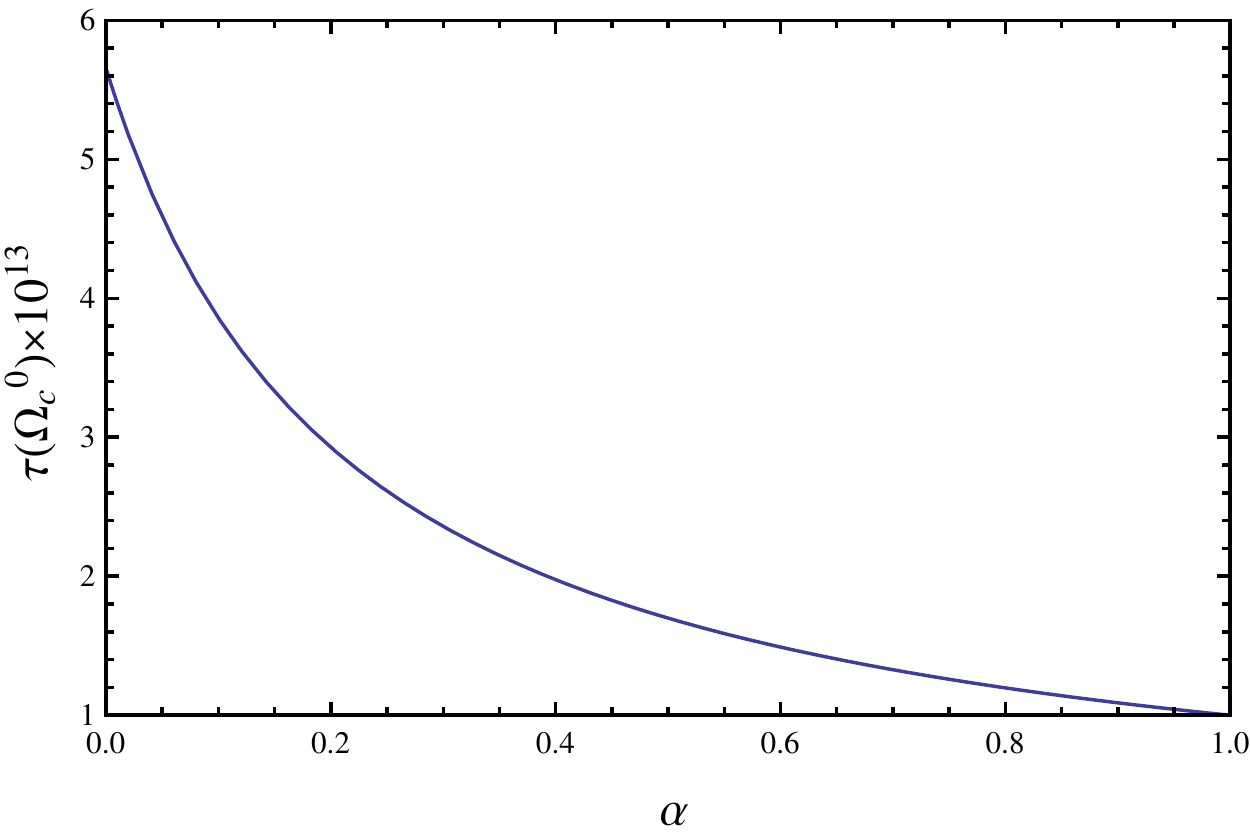}
\caption{Lifetime of the $\Omega_c^0$ as a function of $\alpha$.}
\label{fig:lifetime_Omegac}
\end{center}
\end{figure}

In order to allow a description of the $1/m_c^4$ corrections to $\Gamma(\Omega_c)$ within the realm of perturbation theory, we introduce a  parameter $\alpha$ so that $\Gamma^{\rm int}_{+,7}(\Omega_c)$ and $\Gamma^{\rm SL}_7(\Omega_c)$ are multiplied by a factor of $(1-\alpha)$; that is, $\alpha$ describes the degree of suppression.
The origin of this suppression is unknown, but it could be due to the next-order $1/m_c$ correction.
\footnote{Another possibility is that, as noticed in passing, it is not clear to us what are the baryon matrix elements of dimension-7 operators $P^q_{5,6}$.  This may also explain the suppression needed for $\Gamma^{\rm int}_{+,7}(\Omega_c)$ and $\Gamma^{\rm SL}_7(\Omega_c)$. Note that the above-mentioned suppression effect is not needed for the antitriplet baryons $\Lambda_c^+,\Xi_c^+$ and $\Xi_c^0$.}
The lifetime of $\Omega_c^0$ is plotted as a function of $\alpha$ in Fig. \ref{fig:lifetime_Omegac}. For the two extreme cases that $\alpha=0$ (no suppression on dimension-7 effects) and $\alpha=1$ (no corrections to $\Gamma^{\rm int}_{+}(\Omega_c)$ and $\Gamma^{\rm SL}(\Omega_c)$ from dimension-7 operators),
we have $\tau(\Omega_c^0)=6.10\times 10^{-13}s$ and $0.97\times 10^{-13}s$, respectively (see Table \ref{tab:lifetime4_charmbary_1}).
Our guidelines for the parameter $\alpha$ are (i) both $\Gamma^{\rm int}_{+}(\Omega_c)$ and $\Gamma^{\rm SL}(\Omega_c)$ should be positive (at least, a negative $\Gamma^{\rm SL}(\Omega_c)$ does not make sense), and (ii)
$\Gamma^{\rm SL}(\Omega_c)$ is comparable to that of $\Lambda_c^+$ or $\Xi_c$.
Under these guidelines, we get $\alpha>0.16$ to ensure  positive $\Gamma^{\rm SL}(\Omega_c)$ and $\Gamma^{\rm int}_{+}(\Omega_c)$  and $\alpha\sim 0.22$ (0.32) for $\Gamma^{\rm SL}(\Omega_c)$  to be comparable to that of $\Lambda_c^+$ ($\Xi_c$).

We see from Table \ref{tab:lifetime4_charmbary_1}  that a reasonable range of $\alpha$ lies in $0.16<\alpha<0.32$ and  the corresponding $\Omega_c$ lifetimes lies in the range
\be
2.3\times 10^{-13}s<\tau(\Omega_c^0)<3.3\times 10^{-13}s.
\en
It should be stressed that this is our conjecture as we do not have rigorous statements on the unknown parameter $\alpha$. At any rate, the $\Omega_c^0$ lifetime
is very different from the current world average of $\tau(\Omega_c^0)=(0.69\pm0.12)\times 10^{-13}s$ \cite{PDG} from fixed target experiments. We suggest the new lifetime pattern
\be
\tau(\Xi_c^+)>\tau(\Omega_c^0)>\tau(\Lambda_c^+)>\tau(\Xi_c^0),
\en
which can be tested in the forthcoming LHCb measurements of charmed baryon lifetimes. Very recently, LHCb has reported a new measurement of the $\Omega_c^0$ lifetime, $\tau(\Omega_c^0)=(2.68\pm0.24\pm0.10\pm0.02)\times 10^{-13}s$ \cite{LHCb:Omegac}, using the semileptonic decay $\Omega_b^-\to\Omega_c^0\mu^-\bar \nu_\mu X$ with $\Omega_c^0\to pK^- K^-\pi^+$. This value is nearly four times larger than the current world-average value.
\footnote{
Our early conjecture of $\tau(\Omega_c^0)$ of order $2.3\times 10^{-13}s$ first presented  in \cite{Cheng:HIEPA} is indeed consistent with the LHCb measurement.}

Finally, we would like to remark on the semileptonic widths. We see from Table \ref{tab:lifetime3_charmbary} that to order $1/m_c^3$,  the constructive Pauli interference is sizeable for the $\Xi_c$
and becomes overwhelming for the $\Omega_c$. However, this interference effect will be partially washed out by the next-order $1/m_c$ correction, in particular for the latter (see Table \ref{tab:lifetime4_charmbary_1}).  Nevertheless, the interference effect in semileptonic inclusive decays can be tested by measuring the ratio of semileptonic branching fractions $\B^{\rm semi}(\Xi_c^+)
/\B^{\rm semi}(\Lambda_c^+)$,
where $\B^{\rm semi}(\B_c)=\B(\B_c\to Xe^+\nu_e)$. This ratio naively of order 1.8 will be enhanced to ${\cal O}(3.2)$ in the presence of Pauli interference.

\begin{table}[t]
\caption{Various contributions to the decay rates (in units of
$10^{-12}$ GeV) of the $\Omega_c^0$ after including subleading $1/m_c$ corrections to spectator effects. However, the dimension-7 contributions $\Gamma^{\rm int}_{+,7}(\Omega_c^0)$ and $\Gamma^{\rm SL}_7(\Omega_c^0)$ are multiplied by a factor of $(1-\alpha)$ with $\alpha$ varying from 0 to 1. }
\label{tab:lifetime4_charmbary_1}
\begin{center}
\begin{tabular}{|c c c  r r l  l  |} \hline \hline
$\alpha$ & $\Gamma^{\rm dec}$ & $\Gamma^{\rm ann}$ &
$\Gamma^{\rm int}_+$ & ~ $\Gamma^{\rm semi}$ & ~~$\Gamma^{\rm tot}$ &
~$\tau(10^{-13}s)$~  \\
\hline
 ~0~~ & ~1.019~ & ~0.876~ & $-0.559$ & ~$-0.256$ &
~1.079 & ~ 6.10 \\
 ~0.12~~ & ~1.019~ & ~0.876~ & $-0.135$ & ~$0.003$ &
~1.762 & ~ 3.73 \\
 ~0.16~~ & ~1.019~ & ~0.876~ & $0.006$ & ~$0.089$ &
~1.990 & ~ 3.31 \\
 ~0.22~~ & ~1.019~ & ~0.876~ & $0.218$ & ~$0.219$ &
~2.331 & ~ 2.82 \\
 ~0.32~~ & ~1.019~ & ~0.876~ & $0.571$ & ~$0.435$ &
~2.900 & ~ 2.27 \\
 ~1~~ & 1.019 & 0.876 & 2.974 & ~1.901 &
~6.770 & ~ 0.97  \\
\hline \hline
\end{tabular}
\end{center}
\end{table}

\section{DISCUSSIONS AND CONCLUSIONS}
In this work we have analyzed the lifetimes of bottom and charmed hadrons within the framework of the heavy quark expansion. It is well known that the lifetime differences stem from spectator effects such as weak annihilation and Pauli interference. We list the dimension-6 four-quark operators responsible for various spectator effects and derive the corresponding dimension-7 ones. The hadronic matrix elements of four-quark operators are parameterized in a model-independent way.

\vskip 0.2cm
The main results of our analysis are as follows.
\begin{itemize}

\item 
    Since in this work we focus on the LO-QCD  study for reason of consistency, the inclusive rate to LO is sensitive to the quark mass definition.
    For the $b$ quark mass, we use the kinetic mass $m_b^{\rm kin}=4.546$ GeV obtained from a recent global fit to the inclusive semileptonic $B$ decay to $X_ce^+\nu_e$ in the kinetic scheme.
    Using the dimension-6 bag parameters recently determined from HQET sum rules and the vacuum-insertion approximation for meson matrix elements of dimension-7 operators, the calculated $B$ meson lifetime ratios $\tau(B^+)/\tau(B^0_d)=1.074^{+0.017}_{-0.016}$ and $\tau(B^0_s)/\tau(B^0_d)=0.9962\pm0.0024$ are in excellent agreement with experiment.

\item Baryon matrix elements of four-quark operators parametrized in a model-independent way in terms of four parameters, but only two of them are independent. They are
    evaluated using the NQM and the bag model.  The hadronic parameter $r$ defined in Eq.~(\ref{eq:rb}) is estimated in the NQM to be in the range 0.60 to 0.66 for both bottom and charmed baryons.

\item
The lifetime pattern of  bottom baryons is found to be $\tau
(\Omega_b^-)> \tau(\Xi_b^-)>\tau(\Xi_b^0)\simeq \tau(\Lambda_b^0)$.
Spectator effects due to $W$-exchange and destructive Pauli interference
account for their lifetime differences. The calculated lifetime ratios
$\tau(\Xi_b^-)/\tau(\Lambda_b^0)$, $\tau(\Xi_b^-)/\tau(\Xi_b^0)$ and $\tau(\Omega_b^-)/\tau(\Xi_b^-)$  agree well with the data.
Moreover, the $\Lambda_b-B$ lifetime ratio $\tau(\Lambda_b^0)/\tau(B_d^0)=0.953$ is in good agreement with the experimental average, indicating that the heavy quark expansion in $1/m_b$ works well for bottom hadrons.

\item It is found that $W$-annihilation contribution in $B$ decays is much smaller than that in bottom baryon decays (see Tables \ref{tab:Blifetime} and \ref{tab:bottombaryon}). The $W$-exchange in $B$ decays is helicity suppressed, while it is neither helicity nor color suppressed in the heavy baryon case.

\item Contrary to the bottom hadron sector where the HQE in $1/m_b$ works well, the HQE to $1/m_c^3$ fails to give a satisfactory description of the lifetimes of both charmed mesons and charmed baryons. This calls for the subleading $1/m_Q$ corrections to spectator effects.

\item  We have employed the
experimental values for $D^+$ and $D^0$ semileptonic widths to fix
the charmed quark mass to be $m_c=1.56$ GeV. For the charmed meson decay constant of order 200 MeV, the destructive Pauli interference leads to a negative  $D^+$ width irrespective of which charmed quark mass is employed.
We showed that $1/m_c$ corrections to the Pauli interference arising from dimension-7 four-quark operators will be able to render $\Gamma(D^+)$ positive. We use the measured lifetime ratios of $\tau(D^+)/\tau(D^0)$ and $\tau(D_s^+)/\tau(D^0)$ to constrain the bag parameters and find that $B_2>B_1$.

\item The HQE to order $1/m_c^3$ implies the lifetime hierarchy $\tau(\Xi_c^+)>\tau(\Lambda_c)>\tau(\Xi_c^0)
>\tau(\Omega_c)$.   However, the
quantitative estimates of charmed baryon lifetimes and their ratios are
still rather poor. For example, the large ratios of
$\tau(\Xi_c^+)/\tau(\Lambda_c^+)$ and $\tau(\Xi_c^+)/\tau(\Xi_c^0)$
are not quantitatively understandable.

\item  The calculated lifetimes for heavy baryons depend on the low normalization point. We considered the hadronic scale in the range of $\mu_{\rm had}\sim 0.65-1$ GeV  and found that the lifetimes of bottom baryons stay almost constant with variation of $\mu_{\rm had}$, while the charmed baryon lifetimes increase by around $10\%$ when the hadronic scale varies from 0.65 to 1.0 GeV. Nevertheless, the charmed baryon lifetime ratios remain nearly constant.

\item The relevant dimension-7 spectator effects are in the right direction for explaining the large lifetime ratio of $\tau(\Xi_c^+)/\tau(\Lambda_c^+)$, which is enhanced from 1.05 to 1.88, in better agreement with experiment.
    However, the destructive $1/m_c$ corrections to $\Gamma(\Omega_c^0)$ are too large to justify the use of the HQE, namely, the predicted Pauli interference and semileptonic rates for $\Omega_c^0$ become negative. Demanding these rates to be positive for a sensible HQE, we conjecture that the $\Omega_c^0$ lifetime lies in the range of $(2.3\sim3.2)\times 10^{-13}s$. This leads to the new lifetime pattern
$\tau(\Xi_c^+)>\tau(\Omega_c^0)>\tau(\Lambda_c^+)>\tau(\Xi_c^0)$, contrary to the current hierarchy $\tau(\Xi_c^+)>\tau(\Lambda_c^+)>\tau(\Xi_c^0)>\tau(\Omega_c^0)$. This new charmed baryon lifetime pattern can be tested by LHCb.

\item The $\Omega_c^0$, which is naively expected to be shortest-lived in the charmed baryon system owing to the large constructive Pauli interference, could live longer than the $\Lambda_c^+$  due to the suppression from $1/m_c$ corrections arising from dimension-7 four-quark operators.

\item
For
charmed baryons $\Xi_c$ and $\Omega_c$, there is an additional contribution
to the semileptonic width coming from the constructive Pauli interference
of the $s$ quark. However, this interference effect will be partially washed out by the next-order $1/m_c$ correction, in particular for the latter.  Nevertheless, this interference effect can be tested by measuring the ratio of semileptonic branching fractions $\B^{\rm semi}(\Xi_c^+)
/\B^{\rm semi}(\Lambda_c^+)$.

\end{itemize}

Finally, we would like to remark that it is straightforward to generalize the present lifetime analysis of singly heavy baryons to doubly heavy ones. Recently, LHCb has presented the first measurement of the lifetime of the doubly charmed baryon $\Xi_{cc}^{++}$ to be
$\tau(\Xi_{cc}^{++})=
(2.56^{+0.24}_{-0.22}\pm0.14)\times 10^{-13}s$ \cite{LHCb:Xiccpp}.

\section{Acknowledgments}
 We would like to thank Alexander Lenz, Xiao-Rui Lyu and Thomas Rauh for helpful discussions. This research was supported in part by the Ministry of Science and Technology of R.O.C. under Grant No. 106-2112-M-001-015.

\appendix

\section{Spectator effects from dimension-7 four-quark operators}

In this appendix we sketch the derivation of dimension-7 operators (\ref{eq:T7baryon_original}) relevant for the spectator effects in heavy baryon decays. The term $\T^{\B_Q,q_1}_{7,ann}$ in Eq. (\ref{eq:T7baryon_original}) corresponds to ${\cal T}_4^{\rm PI}$ in Eq. (17) of \cite{Lenz:D}. Consider the Cabibbo-allowed charmed baryon decay so that
\be
\T^{\B_c,d}_{7,ann}={G_F^2m_c^2\over 6\pi}|V_{cs}V_{ud}|^2\sum^6_{i=1}\left(g_i^{su}P_i^d+h_i^{su}S^d_i\right),
\en
with
\footnote{Our convention for the Wilson coefficients $c_1$ and $c_2$ is opposite to that of \cite{Lenz:D}, namely their $C_2$ is our $c_1$ and vice versa.}
\be
g_i^{qq'} &=& c_2^2 g_{i,11}^{qq'}+c_1c_2 g_{i,12}^{qq'}+c_1^2 g_{i,22}^{qq'}, \non \\
h_i^{qq'} &=& c_2^2 h_{i,11}^{qq'}+c_1c_2 h_{i,12}^{qq'}+c_1^2 h_{i,22}^{qq'}.
\en
Since (see Eq. (B3) of \cite{Lenz:D})
\be
g_{1,ij}^{su}=g_{2,ij}^{su}=g_{4,ij}^{su}=h_{1,ij}^{su}=h_{2,ij}^{su}=h_{4,ij}^{su}=0,
\en
it follows that
\be
\T^{\B_c,d}_{7,ann}={G_F^2m_c^2\over 6\pi}|V_{cs}V_{ud}|^2\left(g_3^{su}P_3^d+g_5^{su}P_5^d+g_6^{su}P_6^d+h_3^{su}S^d_3
+h_5^{su}S_5^d+h_6^{su}S_6^d\right).
\en
The coefficients $g_3^{su}$ and $h_3^{su}$ are given by
\be
g_3^{su}=2(1-x^2)(c_1^2+6c_1c_2+c_2^2), \qquad h_3^{su}=12(1-x^2)(c_1^2+c_2^2).
\en
The remaining coefficients $g_{5(6),ij}^{sq}$ and $h_{5(6),ij}^{sq}$ in HQET are related to those in QCD via (see Eq. (B5) of \cite{Lenz:D})
\be
(g_{5,ij}^{sq})_{\rm HQET}=(F_{ij}^{sq})_{\rm QCD}, \qquad (g_{6,ij}^{sq})_{\rm HQET}=(F_{S,ij}^{sq})_{\rm QCD}, \non \\
(h_{5,ij}^{sq})_{\rm HQET}=(G_{ij}^{sq})_{\rm QCD}, \qquad (h_{6,ij}^{sq})_{\rm HQET}=(G_{S,ij}^{sq})_{\rm QCD},
\en
with the coefficients $F^{sq}$ and $G^{sq}$ given in
\be \label{eq:T6QCD}
 \T^{\B_c,d}_{6,ann}={G_F^2m_c^2\over 6\pi}|V_{cs}V_{ud}|^2 \left(F^{su}Q^d+F^{su}_S Q^q_S+ G^{su}T^d+G^{su}_S T^d_S\right),
\en
where
\be
Q^d=(\bar cd)(\bar dc),  \qquad && Q^d_S=\bar c(1-\gamma_5)d\bar d(1+\gamma_5)c, \non \\
T^d=(\bar ct^ad)(\bar dt^a c),  \qquad && T^d_S=\bar c(1-\gamma_5)t^a d\bar d(1+\gamma_5)t^a c.
\en
Identifing Eq. (\ref{eq:T6QCD})  with $\T^{\B_Q,q_1}_{6,ann}$ in Eq. (\ref{eq:T6baryon}), we see that
\be
F^{su}=(1-x)^2(c_1^2+c_2^2+6c_1c_2), \qquad G^{su}=6(1-x)^2(c_1^2+c_2^2), \qquad F_S^{su}=G_S^{su}=0.
\en
Hence,
\be
g_5^{su}=(1-x)^2(c_1^2+c_2^2+6c_1c_2), \qquad h_5^{su}=6(1-x)^2(c_1^2+c_2^2), \qquad g_6^{su}=h_6^{su}=0.
\en
The expression of $\T^{\B_Q,q_1}_{7,ann}$ given in Eq. (\ref{eq:T7baryon_original}) with $\B_Q=\B_c$ and $q_1=d$  is thus obtained.

Likewise, $\T^{\B_c,u}_{7,int}$ given in Eq. (\ref{eq:T7baryon_original}) corresponds to the transition operator $\T^{{\rm WA}_0}_4$ in \cite{Lenz:D} and it has the expression
\be
\T^{\B_c,u}_{7,int}={G_F^2m_c^2\over 6\pi}|V_{cs}V_{ud}|^2\sum^6_{i=1}\left(g_i^{sd}P_i^u+h_i^{sd}S^u_i\right).
\en
From Appendix B of \cite{Lenz:D} we obtain
\be
 && g_1^{sd}=g_2^{sd}=-(1-x)^2(1+2x)\left({1\over N_c}c_1^2+2c_1c_2+N_cc_2^2\right), \non \\
&& g_3^{sd}=2(1-x)(1+x+x^2)\left({1\over N_c}c_1^2+2c_1c_2+N_cc_2^2\right),  \non  \\
&& g_4^{sd}=-12x^2(1-x)\left({1\over N_c}c_1^2+2c_1c_2+N_cc_2^2\right),   \\
 && h_1^{sd}=h_2^{sd}=-2(1-x)^2(1+2x)c_1^2, \non  \\
&& h_3^{sd}=4(1-x)(1+x+x^2)c_1^2,  \qquad h_4^{sd}=-24x^2(1-x)c_1^2.  \non
\en
The coefficients $g_{5,6}^{sd}$ and $h_{5,6}^{sd}$ are found by comparing
\be
\T^{\B_c,u}_{6,int}={G_F^2m_c^2\over 6\pi}|V_{cs}V_{ud}|^2 \left(F^{sd}Q^u+F^{sd}_S Q^u_S+ G^{sd}T^u+G^{sd}_S T^u_S\right)
\en
with $\T^{\B_Q,q_2}_{6,int-}$ in Eq. (\ref{eq:T6baryon}). Hence,
\be
g_5^{sd} &=& -(1-x)^2(1+{x\over 2})\left({1\over N_c}c_1^2+2c_1c_2+N_cc_2^2\right),  \non \\
g_6^{sd} &=& (1-x)^2(1+2x)\left({1\over N_c}c_1^2+2c_1c_2+N_cc_2^2\right),  \non \\
h_5^{sd} &=& -2(1-x)^2(1+{x\over 2})c_1^2, \non \\
h_6^{sd} &=& -2(1-x)^2(1+2x)c_1^2.
\en
This completes the derivation of $\T^{\B_c,u}_{7,int}$.

The $\T^{\B_b,s}_{7,int}$ term in Eq. (\ref{eq:T7baryon_original}) describes the Pauli interference in $b\to c\bar cs$ (see Fig. \ref{fig:spectator}.(b)). It can be deduced from
\be
\T^{\B_b,s}_{7,int}={G_F^2m_b^2\over 6\pi}|V_{cb}V_{cs}|^2\sum^6_{i=1}\left(g_i^{ss}P_i^s+h_i^{ss}S^s_i\right).
\en
We find
\be
&& g_1^{ss}=g_2^{ss}=-\sqrt{1-4x}\,(1+2x)\left({1\over N_c}c_1^2+2c_1c_2+N_cc_2^2\right),  \non \\
&& g_3^{ss}={2(1-2x-2x^2)\over\sqrt{1-4x}}\left({1\over N_c}c_1^2+2c_1c_2+N_cc_2^2\right), \non \\
&& g_4^{ss}=-{24x^2\over\sqrt{1-4x}}\left({1\over N_c}c_1^2+2c_1c_2+N_cc_2^2\right),  \non \\
&& h_1^{ss}=h_2^{ss}=-2\sqrt{1-4x}\,(1+2x)c_1^2,  \non \\
&& h_3^{ss}={4(1-2x-2x^2)\over\sqrt{1-4x}}c_1^2,  \qquad h_4^{ss}=-{48x^2\over\sqrt{1-4x}}c_1^2,
\en
and
\be
g_5^{ss}&=& -\sqrt{1-4x}\,(1-x)\left({1\over N_c}c_1^2+2c_1c_2+N_cc_2^2\right), \non \\
g_6^{ss}&=& \sqrt{1-4x}\,(1+2x)\left({1\over N_c}c_1^2+2c_1c_2+N_cc_2^2\right), \non \\
h_5^{ss} &=& -2\sqrt{1-4x}\,(1-x)c_1^2, \qquad h_6^{ss} = 2\sqrt{1-4x}\,(1+2x)c_1^2.
\en

Finally, the last term $\T^{\B_c,s}_{7,int}$ in  Eq. (\ref{eq:T7baryon_original}) corresponds to the transition operator $\T_4^{{\rm WA}_+}$ in \cite{Lenz:D}
\be
\T^{\B_c,s}_{7,int}={G_F^2m_c^2\over 6\pi}|V_{cs}V_{ud}|^2\sum^6_{i=1}\left(\tilde g_i^{ud}P_i^s+\tilde h_i^{ud}S^s_i\right).
\en
It turns out that it has the same expression as  $\T^{\B_Q,q_3}_{7,int}$ except for a vanishing $x$ and the interchange of $c_1$ and $c_2$.

\section{Baryon matrix elements in the quark model}
We show briefly the derivation of Eqs. (\ref{eq:m.e.baryon}), (\ref{eq:m.e.baryon_1}) and (\ref{eq:m.e.1}) in the MIT bag model because the expressions in the non-relativistic quark model can be obtained from the former through a simple replacement given in Eq. (\ref{eq:replacement}). Consider the four-quark operator $O=(\bar QQ)(\bar qq)$. This operator can be written as
$O=6(\bar QQ)_1(\bar qq)_2$, where the superscript $i$ indicates that the quark operator acts only on the $i$th quark in the baryon wave function.
In the bag model, it has the expression (see e.g. Eq. (B2) of \cite{CT})
\be
(\bar QQ)_1(\bar qq)_2=a_q+b_q-\left(a_q-{b_q\over 3}+{8c_q\over 3}\right){\vec\sigma}_Q\cdot{\vec\sigma}_q,
\en
where $a_q$, $b_q$ and $c_q$ are the four-quark overlap integrals defined in Eq. (\ref{eq:overlap}) in terms of the large and small components of the quark wave function,  $u(r)$ and $v(r)$, respectively,
\be \label{eq:uv}
\psi=
\left(
\begin{array}{c}
iu(r)\chi \\
v(r){\vec \sigma}\cdot{\bf \hat{r}}\chi \\
\end{array}
\right).
\en
Applying the relation
\be
{\vec\sigma}_1\cdot{\vec\sigma}_2={1\over 2}(\sigma_{1+}\sigma_{2-}+\sigma_{1-}\sigma_{2+})+\sigma_{1z}\sigma_{2z},
\en
and the wave functions
\be
\Lambda_b^0 &=& -{1\over \sqrt{6}}\left[(ud-du)\chi_{_A}+(13)+(23)\right], \non \\
\Omega_b^- &=& {1\over \sqrt{3}}\left[ssb\chi_{_S}+(13)+(23)\right],
\en
with obvious notation for permutation of quarks,
where $abc\chi_{_S}=(2a^\up b^\up c^\dw-a^\up b^\dw c^\up-a^\dw b^\up c^\up)/\sqrt{6}$ and $abc\chi_{_A}=(a^\up b^\dw c^\up-a^\dw b^\up c^\up)/\sqrt{2}$,
it is straightforward to show that
\be
\la \Omega_b|b^\dagger_{1b}b_{1b}b^\dagger_{2s}b_{2s} |\Omega_b\ra={1\over 3}, \qquad
\la \Omega_b|b^\dagger_{1b}b_{1b}b^\dagger_{2s}b_{2s} \, ({\vec\sigma}_{b1}\cdot {\vec\sigma}_{s2}) |\Omega_b\ra=-{2\over 3}.
\en
Hence,
\be
\la \Omega_b|(\bar bb)(\bar ss)|\Omega_b\ra=6 \la \Omega_b^0|(\bar bb)_1(\bar ss)_2|\Omega_b\ra={1\over 3}(18a_s+2b_s+32c_s).
\en
Likewise,
\be
\la \Lambda_b|(\bar bb)(\bar qq)|\Lambda_b\ra=a_q+b_q.
\en

Next, using the expression
\be
(\bar Q\gamma_\mu\gamma_5 Q)_1(\bar q\gamma^\mu(1-\gamma_5)q)_2 &=&(a_q-b_q){\vec\sigma}_Q\cdot{\vec\sigma}_q+2b_q({\vec\sigma}_Q\cdot{\bf \hat r})({\vec\sigma}_q\cdot{\bf \hat r}) \non \\
&=& \left(a_q-{b_q\over 3}\right){\vec\sigma}_Q\cdot{\vec\sigma}_q.
\en
we obtain
\be
\la \Omega_b|\bar b\gamma_\mu\gamma_5 b\bar s\Gamma^\mu s|\Omega_b\ra=-4\left(a_s-{b_s\over 3}\right), \qquad \la \Lambda_b|\bar b\gamma_\mu\gamma_5 b\bar q\Gamma^\mu q|\Lambda_b\ra=0,
\en
where $\Gamma^\mu=\gamma^\mu(1-\gamma_5)$.
With Eq. (\ref{eq:rel}) and the relation
\be
\bar b\gamma_\mu\gamma_5 b\bar q\gamma^\mu(1-\gamma_5)q=-\bar b^\alpha(1-\gamma_5)q^\beta \bar q^\beta(1+\gamma_5)b^\alpha-{1\over 2}(\bar bq)(\bar
q b),
\en
we arrive at the results of (\ref{eq:m.e.baryon_1}) and (\ref{eq:m.e.1}).

Since in heavy quark effective theory, the matrix element $\la H_Q|\bar Qv\!\!\!/ Q|H_Q\ra$ is normalized to $2m_{H_Q}$, we need to put back the factor of $2m_{H_Q}$ in an appropriate place, for example, $\la \Lambda_b^0|(\bar bb)(\bar qq)|\Lambda_b\ra$ now reads $(a_q+b_q)(2m_{\Lambda_b})$. Note that in the quark model, the hadronic parameter $\tilde B$ is equal to unity which is supposed to be valid at the hadronic scale.

%
%
\newcommand{\bi}{\bibitem}
%

\end{document}